\documentclass[aps,twocolumn,pra,superscriptaddress,amsmath,showpacs,tightenlines]{revtex4-1}
\usepackage{amssymb}
\usepackage{amsmath}
\usepackage{graphicx}
\usepackage{subfigure}
\usepackage{natbib}
\usepackage{epsfig}
\usepackage{amsfonts}
\usepackage{mathrsfs}
\usepackage{xcolor}
\usepackage[toc,page,title,titletoc,header]{appendix}

\begin{document}

\title{Steady-state entanglement and coherence of the coupled qubit system in equilibrium and nonequilibrium environments}
\author{Zhihai \surname {Wang}}
\email{Z Wang and W Wu contributed equally to this paper.}
\affiliation{Center for quantum Sciences and School of Physics, Northeast Normal University, Changchun 130024, China}
\author{Wei \surname{Wu}}
\email{Z Wang and W Wu contributed equally to this paper.}
\affiliation{State Key Laboratory of Electroanalytical Chemistry, Changchun Institute of Applied Chemistry, Chinese Academy of Sciences, Changchun 130022, China}
\author{Jin \surname{Wang}}
\email{Email: jin.wang.1@stonybrook.edu}
\affiliation{State Key Laboratory of Electroanalytical Chemistry, Changchun Institute of Applied Chemistry, Chinese Academy of Sciences, Changchun 130022, China}
\affiliation{Department of Chemistry and Department of Physics and Astronomy, State University of New York at Stony Brook, NY 11794, USA}

\begin{abstract}
We analytically and numerically investigate the steady-state entanglement and coherence of two coupled qubits each interacting with a local boson or fermion reservoir, based on the Bloch-Redfield master equation beyond the secular approximation. We find that there is non-vanishing steady-state coherence in the nonequilibrium scenario, which grows monotonically with the nonequilibrium condition quantified by the temperature difference or chemical potential difference of the two baths. The steady-state entanglement, in general, is a non-monotonic function of the nonequilibrium condition as well as the bath parameters in the equilibrium setting. We also discover that weak inter-qubit coupling and high base temperature or chemical potential of the baths can strongly suppress the steady-state entanglement and coherence, regardless of the strength of the nonequilibrium condition. On the other hand, the energy detuning of the two qubits, when used in a compensatory way with the nonequilibrium condition, can lead to significant enhancement of the steady-state entanglement in some parameter regimes. In addition, the qubits typically have a stronger steady-state entanglement when coupled to fermion baths exchanging particles with the system than boson baths exchanging energy with the system, under similar conditions.  {We also identify a close connection between the energy current flowing through the system and the steady-state coherence. Preliminary investigations suggest that these results are insensitive to the form of the reservoir spectral densities in the Markovian regime.} Feasible experimental realization of measuring the steady-state entanglement and coherence is discussed for the coupled qubit system in nonequilibrium environments. Our findings offer some general guidelines for optimizing the steady-state entanglement and coherence in the coupled qubit system and may find potential applications in quantum information technology.

\end{abstract}

\maketitle

\section{Introduction}

Entanglement and coherence are fundamental concepts in quantum mechanics and key resources in quantum information processing~\cite{MA,CH0,CH,BS,scully,VS}. In practice, a quantum system inevitably interacts with its surrounding environments, deteriorating useful quantum resources such as entanglement and coherence very quickly as a result of the decoherence process~\cite{HB,WHZ}. It is thus critical to study the generation, control and protection of entanglement and coherence in the context of open quantum systems. Recently, there has been a growing interest in the study of open quantum systems in nonequilibrium environments \cite{NoneqEnt1,NoneqEnt2,NoneqEnt3,NoneqEnt4,NoneqEnt5,NoneqEnt6,NoneqEnt7}. It has been suggested that entanglement and coherence can survive in nonequilibrium steady states sustained by nonequilibrium environments in constant exchange of matter, energy and information with the quantum system~\cite{NoneqEnt5, NoneqEnt6,NoneqEnt7,en3,en4,en6,en7,en8,en9,coh1,coh2,coh3,coh4,coh5,coh6,coh7,
ExactlySolvable1,ExactlySolvable2,ExactlySolvable3,Fermion2,NoneqFluc,Trasport,ArminTavakoli}. The nonequilibrium conditions in the environments are maintained by a constant temperature difference or chemical potential difference that drives the flow of energy or matter through the quantum system and the environments, sustaining a steady deviation from the thermodynamic equilibrium \cite{ExactlySolvable1,ExactlySolvable2,ExactlySolvable3,ArminTavakoli}. The surviving quantum features in nonequilibrium steady states open new doors to the generation, protection and control of quantum resources in open quantum systems through nonequilibrium conditions.

The simplest model imaginable that allows for the investigation of entanglement and coherence in nonequilibrium steady states is probably a coupled qubit system interacting with two reservoirs. As a simple model the two-qubit system has been investigated in various settings~\cite{NoneqEnt4,NoneqEnt6,NoneqEnt7,NoneqEnt5,en3,en8,en9, JQLiao,Marzolino,Marzolino1,Marzolino2,BCAK,Marko,wang1,wang2,MC,DG, MO,AK,LD,FBenatti,FBenatti1,cao,XZ,JonatanBohr,nonMarkovian2,CQ,CQDriven1,CQDriven2,CQDriven3,Fermion2,secBoseFermi,SecCQ,SecNCQ2}. The concept of thermal entanglement was proposed in a spin chain system at thermal equilibrium~\cite{Marko,wang1,wang2,MC,DG}. The dynamical evolution of the entanglement  was explored for two uncoupled qubits interacting with two reservoirs~\cite{LD,FBenatti,FBenatti1,cao,MO}. The coupled qubit system has also been investigated when time-dependent external driving is present~\cite{CQDriven1,CQDriven2,CQDriven3}. The steady-state entanglement of two coupled qubits interacting with two reservoirs have been extensively studied based on Markovian quantum master equations under the secular approximation ~\cite{NoneqEnt6,NoneqEnt7,en8,en9,JQLiao,CQ,SecNCQ2,SecCQ,NoneqEnt5,Fermion2,
NoneqEnt4,secBoseFermi,Ali}. Some aspects of the non-Markovian effects in the two-qubit system has also been explored~\cite{cao,LD,nonMarkovian2,AK,XZ,JonatanBohr}. In this paper, we analytically and numerically investigate the steady-state entanglement and coherence of the coupled qubit system interacting with two independent boson or fermion reservoirs that can exchange energy (boson reservoir) or particle (fermion reservoir) with the system in both equilibrium and nonequilibrium settings, with an emphasis on the entanglement in the nonequilibrium setting.

We adopt a quantum master equation approach in the framework of the Born-Markov approximation, \emph{without} performing the frequently applied secular approximation. The non-secularized Markovian quantum master equation is usually referred to as the Bloch-Redfield equation \cite{Bloch,Redfield}, which has found wide applications in the study of nuclear magnetic resonance \cite{NMR}, chemical dynamical systems \cite{RedfieldApp2}, quantum transport \cite{RedfieldApp3,RedfieldApp4}, and photosynthetic reactions \cite{RedfieldApp5,RedfieldApp6}. A known issue related to the Bloch-Redfield equation is that it does not guarantee \emph{a priori} the positivity of the density matrix in the time evolution, which has been a subject of debate with a long history  \cite{Spohn,RedfieldApp3,RedfieldApp6}. This point has been used to argue for secularizing the Bloch-Redfield equation in favor of the Lindblad equation that is completely positive. However, the secularized master equation ignores important effects such as nonequilibrium steady-state coherence~\cite{coh1,coh2,coh3,coh4,coh5,coh6,coh7}. Moreover, the validity of secularization in certain situations has been questioned as the procedure may lead to physically inconsistent results, such as disregarding the nonequilibrium flux inside a composite system \cite{coh2} and violation of conservation laws \cite{Conservation}. A partial secular approximation scheme has been proposed to limit the indiscriminate use of the secularization procedure \cite{RedfieldApp6,partialsecular}. Furthermore, some studies have suggested that positivity of the density matrix in the Bloch-Redfield equation can be restored without the secular approximation, provided that the initial conditions are restricted to those compatible with the system-bath correlations \cite{InitialCondition}, a consistent noise model for the bath is used \cite{RedfieldApp6}, and the Markovian approximation is truly honored \cite{CSBR}. Basically, violation of positivity is an indication that the equation has been applied outside its range of validity, and thus positivity may be guaranteed by operating inside its validity regimes. However, it is in general a highly non-trivial task to \emph{quantify} the validity regime of a master equation. For an exactly solvable model of boson modes coupled to boson baths \cite{ExactlySolvable1,ExactlySolvable2,ExactlySolvable3}, certain aspects of the validity regime of Markovian master equations have been investigated \cite{ValidityRegime2,CSBR}. Yet much more work is still needed before this issue can be fully understood and resolved. In this study we employ the non-secularized Bloch-Redfield equation and exercise caution in working within parameter regimes ensuring the positivity of the density matrix. We make comments if the issue of violation of positivity arises. A full quantification of the validity regime of the equation is reserved for future work.

We analytically solve the steady state of the Bloch-Redfield equation for the coupled qubit system in the general nonequilibrium setting for both boson and fermion baths, even when the two qubits have an energy detuning. The analytical solution offers insights on the behaviors of the steady-state entanglement and coherence in some extreme parameter regimes, which can be extrapolated to account for their features in moderate parameter regimes where the analytical solution may not be so apt to generate insights. Numerical results are also used in this work as a consistency check for the analytical solution and to explore wider parameter regimes that are difficult to access directly from the analytical solution. Combined with the perspective that the concurrence quantifying the entanglement between the coupled qubits can be interpreted as a competition between coherence and population in the bare-state representation, numerical results can provide another view on some features of the steady-state entanglement.

We investigate the steady-state entanglement and coherence first for the equilibrium  setup and then move on to the nonequilibrium scenario. Within each setting we study the boson and fermion bath case, respectively. The steady-state entanglement in relation to the detuning of the two qubits and the nonequilibrium condition are studied in the entanglement phase diagrams. {We also present some preliminary results on the effect of spectral densities and the connection to energy current.} Generally speaking, we find that the steady-state coherence (in the eigen-state representation) has a simpler behavior, while the steady-state entanglement displays more complicated features. More specifically, the steady-state coherence vanishes in the equilibrium scenario as a result of decoherence, and grows monotonically with the nonequilibrium condition characterized by the temperature difference or chemical potential difference. This nonequilibrium steady-state coherence would have been ignored by the secularized quantum master equation. On the other hand, the steady-state entanglement in general varies non-monotonically with the bath parameters (temperatures or chemical potentials as well as their differences) in both equilibrium and nonequilibrium settings. Too weak inter-qubit coupling strength and too high base temperature or chemical potential of the baths both have destructive effects on the steady-state entanglement and coherence, regardless of the strength of the nonequilibrium condition. Combining the detuning of the two qubits with the nonequilibrium condition in a compensatory way (i.e., the qubit with a higher frequency is coupled to the bath with a lower temperature or chemical potential) can lead to significant improvement in the steady-state entanglement ($5\sim 10$ times in some parameter regimes) compared to the equilibrium symmetric qubit case. In addition, fermion baths that exchange particles with the system typically has a beneficial effect on entanglement enhancement, in comparison with boson baths that exchange energy with the system. {We also observe a close connection between the energy current  and the coherence at the steady state. Preliminary investigations suggest that, in the Markovian regime, our results are not sensitive to the form of the spectral densities of the reservoirs.} We also discuss feasible experimental realization of measurements of the steady-state entanglement and coherence for nonequilibrium two-qubit systems. These results provide some general guidelines for enhancing the steady-state entanglement and coherence in the coupled qubit system, which may have potential applications in quantum information processing.

The rest of the paper is organized as follows. In Sec.~\ref{model}, we describe the model and derive the quantum master equation beyond the secular approximation. In Sec.~\ref{ent}, we provide an interpretation of the concurrence that quantifies entanglement as a competition between coherence and population in the bare-state representation. The steady-state entanglement and coherence for the equilibrium and nonequilibrium scenarios are studied in Sec.~\ref{equili} and Sec.~\ref{nonequili}, respectively, for symmetric qubits without energy detuning. In Sec.~\ref{phased}, we discuss the analytical solution and the entanglement phase diagrams for asymmetric qubits with an energy detuning. {Some preliminary investigations on the effect of spectral densities and the connection to energy current are presented in Sec.~\ref{SpectralEnergy}.} The conclusion is summarized in Sec.~\ref{conclusion}. The dynamical equations for the density matrix elements and the method of solving analytically the steady state of the Bloch-Redfield equation are given in the Appendix.

\section{Model and Master equation}
\label{model}
\begin{figure}[tbp]
\centering
\includegraphics[width=8cm]{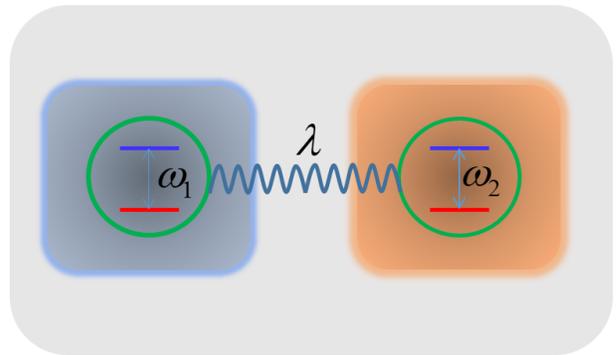}
\caption{(Color online) Schematic diagram of the physical model under consideration. The coupled qubit system, with respective energy spacings $\omega_1$ and $\omega_2$, are coupled to each other and immersed in their individual reservoirs.
}
\label{scheme}
\end{figure}

The model under consideration is illustrated in Fig.~\ref{scheme}. Two qubits (or two-level systems) are coupled to each other, and each qubit is embedded in its own reservoir that follows either bosonic or fermionic statistics. The Hamiltonian for the total system reads $H=H_s+H_R+V$, where ($\hbar=k_B=1$ in the following)
\begin{eqnarray}
H_{s}&=&\omega_{1}|e\rangle_{1}\langle e|+\omega_{2}|e\rangle_{2}\langle e|+\frac{\lambda}{2}[\sigma_{+}^{(1)}\sigma_{-}^{(2)}+\sigma_{-}^{(1)}\sigma_{+}^{(2)}],
\\H_{R}&=&\sum_{k}\omega_{bk}b_{k}^{\dagger}b_{k}+\sum_{k}\omega_{ck}c_{k}^{\dagger}c_{k},
\\V&=&\sum_{k}g_{k}[\sigma_{-}^{(1)}b_{k}^{\dagger}+\sigma_{+}^{(1)}b_{k}]+\sum_{k}f_{k}
[\sigma_{-}^{(2)}c_{k}^{\dagger}+\sigma_{+}^{(2)}c_{k}].\nonumber \\
\label{Hamiltonian}
\end{eqnarray}
$H_s$ is the Hamiltonian of the coupled qubit system, where $\omega_1$ and $\omega_2$ are the respective energy-level spacings (frequencies) of the two qubits and $\lambda$ is the inter-qubit coupling strength. $H_R$ is the free Hamiltonian of the reservoirs, where $b_k$ ($b_k^{\dagger}$) and $c_k$ ($c_k^{\dagger}$) are the annihilation (creation) operators for the $k$-th mode with frequencies $\omega_{ck}$ and $\omega_{dk}$ in the reservoirs in contact with qubit $1$ and $2$, respectively. The last term $V$ is the qubit-reservoir interaction Hamiltonian under the rotating wave approximation, and $g_k$ and $f_k$ are the qubit-reservoir coupling strengths assumed to be real.

\begin{figure}[tbp]
\centering
\includegraphics[width=8cm]{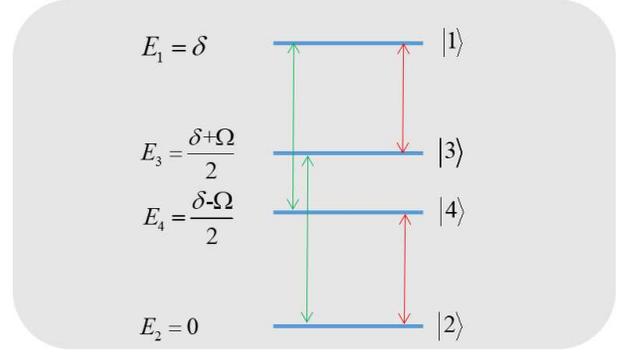}
\caption{(Color online) The levels of the four eigenstates $|E_i\rangle$ for $i=1,2,3,4$ of the coupled qubit system. The red and green lines represent two groups of energy-level transitions induced by the system-reservoir interaction (see the text for detail).}
\label{transition}
\end{figure}

The eigen-energies and the corresponding eigen-states of the Hamiltonian for the
coupled qubit system $H_s$ are obtained as follows~\cite{en9}:
\begin{subequations}
\begin{eqnarray}
E_{1}&=&\delta,\,\,\,\,\,\,\,\,\,\,\,\,\,\,|1\rangle=|ee\rangle,\\
E_{2}&=&0,\,\,\,\,\,\,\,\,\,\,\,\,\,\,|2\rangle=|gg\rangle,\\
E_{3}&=&	\frac{\delta+\Omega}{2},\,|3\rangle=\cos\frac{\theta}{2}|eg\rangle+
\sin\frac{\theta}{2}|ge\rangle,\\
E_{4}&=&	\frac{\delta-\Omega}{2},\,|4\rangle=
-\sin\frac{\theta}{2}|eg\rangle+\cos\frac{\theta}{2}|ge\rangle,
\end{eqnarray}
\end{subequations}
where $\delta=\omega_1+\omega_2$, $\Delta=\omega_1-\omega_2$, $\Omega=\sqrt{\Delta^{2}+\lambda^{2}}$ is the Rabi frequency, and $\theta\in[0,\pi]$ is the mixing angle defined by $\tan\theta=\lambda/\Delta$. In the symmetric qubit case $\Delta=\omega_1-\omega_2=0$, we have $\theta=\pi/2$. For the asymmetric qubit case, $\theta=\arctan(\lambda/\Delta)$ when $\omega_1>\omega_2$ and $\theta=\pi+\arctan(\lambda/\Delta)$ when $\omega_1<\omega_2$. Notice that to guarantee the validity of the rotating wave approximation in the qubit-reservoir interaction Hamiltonian, it is required that $\lambda<2\sqrt{\omega_1\omega_2}$, which implies $\delta>\Omega$ and thus the eigen-energies form an ordered sequence $E_1>E_3>E_4>E_2$. A schematic representation of the eigen-energies and eigen-states is shown in Fig.~\ref{transition}. The unitary transformation matrix $U$ between the eigen basis $\{|1\rangle,|2\rangle,|3\rangle,|4\rangle\}$ and the bare basis $\{|ee\rangle,|gg\rangle,|eg\rangle,|ge\rangle\}$, defined by $U_{ai}=\langle a|i\rangle$ ($a$ and $i$ label the bare and eigen states respectively), has the explicit expression
\begin{equation}
U=\left(\begin{array}{cccc}
1 & 0 & 0 & 0\\
0 & 1 & 0 & 0\\
0 & 0 & \cos\frac{\theta}{2} & -\sin\frac{\theta}{2}\\
0 & 0 & \sin\frac{\theta}{2} & \cos\frac{\theta}{2}
\end{array}\right).
\end{equation}

In terms of the eigen basis, the interaction Hamiltonian $V$ can be re-expressed
as
\begin{eqnarray}\label{V1}
V&=&\sum_{k}[g_{k}(A_{1}+B_{1})b_{k}^{\dagger} +f_{k}(A_{2}+B_{2})c_{k}^{\dagger}]+h.c.
\end{eqnarray}
where
\begin{subequations}
\begin{eqnarray}\label{AB}
A_{1}&=&\sin\frac{\theta}{2}(|3\rangle\langle1|-|2\rangle\langle4|),\\ B_{1}&=&\cos\frac{\theta}{2}(|4\rangle\langle1|+|2\rangle\langle3|),\\
A_{2}&=&\cos\frac{\theta}{2}(|3\rangle\langle1|+|2\rangle\langle4|),\\ B_{2}&=&\sin\frac{\theta}{2}(|2\rangle\langle3|-|4\rangle\langle1|).
\end{eqnarray}
\end{subequations}
In the interaction picture, with free Hamiltonian $H_0=H_s+H_R$, we have
\begin{eqnarray}\label{Vt}
V(t)&=&	\sum_{k}g_{k}[A_{1}e^{-i\frac{\delta-\Omega}{2}t}
+B_{1}e^{-i\frac{\delta+\Omega}{2}t}]b_{k}^{\dagger}e^{i\omega_{bk}t}+h.c.\nonumber\\
&+&\sum_{k}f_{k}[A_{2}e^{-i\frac{\delta-\Omega}{2}t}+B_{2}
e^{-i\frac{\delta+\Omega}{2}t}]c_{k}^{\dagger}e^{i\omega_{ck}t}+h.c..\nonumber \\
\end{eqnarray}
Under the Born-Markov approximation, the quantum master equation in the interaction picture reads~\cite{HB}
\begin{equation}
\frac{d\rho_I}{dt}=-\int_{0}^{\infty}ds{\rm Tr}_{B}[V(t),[V(t-s),\rho_I(t)\otimes\rho_B]],
\end{equation}
where $\rho_I$ is the reduced density operator of the coupled qubit system in the interaction picture, $\rho_B=\rho_{B1}^{eq}\otimes \rho_{B2}^{eq}$ is the density operator of the reservoirs with each reservoir at its own equilibrium state, and $\rm{Tr}_B$ denotes the partial trace with respect to the degrees of freedom of the reservoirs.

Going back to the Schr\"odinger picture, \emph{without} making the secular approximation~\cite{coh2,coh3,coh6}, we finally arrive at the quantum master equation for the reduced density operator of the system, namely, the Bloch-Redfield equation:
\begin{equation}
\frac{d\rho}{dt}=-i[H_s,\rho]+D_0[\rho]+D_s[\rho],\label{master}
\end{equation}
where
\begin{equation}
D_0[\rho]=\sum_{i=1}^{2}\mathcal{N}_i[\rho],\,D_s[\rho]=\sum_{i=1}^{2}\mathcal{S}_i[\rho],\label{dissipator}
\end{equation}
 and
\begin{eqnarray}
\mathcal{N}_i[\rho]&=&\gamma_{i}^{+}[2B_{i}^{\dagger}\rho B_{i}-B_{i}B_{i}^{\dagger}\rho-\rho B_{i}B_{i}^{\dagger}]\nonumber\\
&+&\gamma_{i}^{-}[2A_{i}^{\dagger}\rho A_{i}-A_{i}A_{i}^{\dagger}\rho-\rho A_{i}A_{i}^{\dagger}]\nonumber\\
&+&
\Gamma_{i}^{+}[2B_{i}\rho B_{i}^{\dagger}-B_{i}^{\dagger}B_{i}\rho-
\rho B_{i}^{\dagger}B_{i}]\nonumber \\
&+&
\Gamma_{i}^{-}[2A_{i}\rho A_{i}^{\dagger}-A_{i}^{\dagger}A_{i}\rho-
\rho A_{i}^{\dagger}A_{i}],\label{seque}\\
\mathcal{S}_i[\rho]&=&\gamma_{i}^{+}[A_{i}^{\dagger}\rho B_{i}+B_{i}^{\dagger}\rho A_{i}-A_{i}B_{i}^{\dagger}\rho-\rho B_{i}A_{i}^{\dagger}]\nonumber\\
&+&\gamma_{i}^{-}[A_{i}^{\dagger}\rho B_{i}+B_{i}^{\dagger}\rho A_{i}-B_{i}A_{i}^{\dagger}\rho-\rho A_{i}B_{i}^{\dagger}]\nonumber \\
&+&\Gamma_{i}^{+}[A_{i}\rho B_{i}^{\dagger}+B_{i}\rho A_{i}^{\dagger}-A_{i}^{\dagger}B_{i}\rho-\rho B_{i}^{\dagger}A_{i}]\nonumber \\
&+&\Gamma_{i}^{-}[A_{i}\rho B_{i}^{\dagger}+B_{i}\rho A_{i}^{\dagger}-B_{i}^{\dagger}A_{i}\rho-\rho A_{i}^{\dagger}B_{i}].\label{noseque}
\end{eqnarray}
In the above, $\gamma_i^{\pm}$ and $\Gamma_i^{\pm}$ are short notations for $\gamma_i(\delta/2\pm\Omega/2)$ and $\Gamma_i(\delta/2\pm\Omega/2)$, respectively.
For boson reservoirs,
\begin{equation}\label{BosonCoefficients}
\gamma_{i}(\omega)=J_{i}(\omega)N_{i}(\omega),\,
\Gamma_{i}(\omega)=J_{i}(\omega)[N_{i}(\omega)+1],
\end{equation}
 and for fermion reservoirs
 \begin{equation}\label{FermionCoefficients}
\gamma_{i}(\omega)=J_{i}(\omega)N_{i}(\omega),\,
\Gamma_{i}(\omega)=J_{i}(\omega)[1-N_{i}(\omega)].
\end{equation}
Here $J_{1}(\omega)=\pi\sum_{k}g_{k}^{2}\delta(\omega-\omega_{bk})$ and
$J_{2}(\omega)=\pi\sum_{k}f_{k}^{2}\delta(\omega-\omega_{ck})$ are the spectral densities of the two reservoirs in contact with qubit $1$ and $2$, respectively~\cite{legget}. $N_i(\omega)=\{\exp[(\omega-\mu_i)/T_{i}]\mp 1\}^{-1}$ is the average particle number on frequency $\omega$ in the $i$-th reservoir, which follows the Bose-Einstein statistics (minus sign) for boson reservoirs and Fermi-Dirac statistics (plus sign) for fermion reservoirs, with $\mu_i$ and $T_i$ the chemical potential and the temperature of the $i$-th reservoir, respectively. For boson reservoirs encountered in practice (such as photon or phonon baths), it is typical that the particle number is not conserved and as a result the chemical potential vanishes. Therefore, unless explicitly stated, we shall set $\mu_i=0$ ($i=1, 2$) for boson reservoirs, namely, $N_i(\omega)=[\exp(\omega/T_{i})- 1]^{-1}$. For fermion reservoirs we retain the chemical potentials, which means the system can exchange particles with the fermion reservoirs in processes conserving the particle number (e.g., in quantum dot systems). Note that in deriving the quantum master equation in Eq.~(\ref{master}), we have neglected the frequency shift terms. The quantum master equation expressed in terms of the density matrix elements is given in Appendix \ref{appendix1}. {In the following sections of this paper, with the exception of Sec.~\ref{OhmicSpectrum}, we restrict ourselves to balanced and frequency-independent spectral densities, $J_1(\delta/2\pm\Omega/2)=J_2(\delta/2\pm\Omega/2)=J$. The effect of unbalanced and frequency-dependent spectral densities (Ohmic spectrum) is discussed in Sec.~\ref{OhmicSpectrum}.}

To better understand the physical processes described by the quantum master equation, we notice that the form of the interaction Hamiltonian $V$ in Eqs.~(\ref{V1})-(\ref{Vt}) reveals that the interaction between the coupled qubit system and the reservoirs induces two groups of energy-level transitions in the system as shown in Fig.~\ref{transition}. One group consists of the transitions $|2\rangle\leftrightarrow|4\rangle$ and $|3\rangle\leftrightarrow|1\rangle$ with the frequency $(\delta-\Omega)/2$, which is denoted by the red arrows. The other group consists of the transitions $|1\rangle\leftrightarrow|4\rangle$ and $|3\rangle\leftrightarrow|2\rangle$ with the frequency $(\delta+\Omega)/2$, denoted by the green arrows. Now we can better appreciate the physical meaning of the dissipators in Eqs.~(\ref{seque})-(\ref{noseque}). The dissipator $D_0[\rho]$ describes processes in which the energy emitted to the reservoirs when the system undergoes an energy-level transition is re-absorbed by the transitions in the same group with the same frequency. In contrast, the dissipator $D_s[\rho]$ describes processes in which the emission and re-absorption of energy are associated with energy-level transitions in different groups with different frequencies. Since the energy level transition frequencies in the two groups are in general different, the process described by $D_s[\rho]$ is usually considered a fast oscillating process and discarded by performing the so-called secular approximation.

The secular approximation works reasonably well in the equilibrium situation, with $T_1=T_2$ for boson reservoirs or $T_1=T_2,\mu_1=\mu_2$ for fermion reservoirs. In such a situation, the diagonal elements of the density matrix are decoupled from the off-diagonal ones, as can be checked from the dynamical equation for the density matrix elements in Appendix \ref{appendix1}.  As a result, the density matrix of the equilibrium steady state is a diagonal matrix, without any coherence left in the energy eigen-state representation. However, in the nonequilibrium situation, with $T_1\neq T_2$ for boson reservoirs and $\mu_1\neq \mu_2$ and/or $T_1\neq T_2$ for fermion reservoirs, the secular approximation ignores important effects, such as the quantum coherence in the nonequilibrium steady state~\cite{coh2,coh3,coh6}. Since nonequilibrium steady-state coherence is an important aspect in our study, we retain the non-secular terms $D_s[\rho]$ in the quantum master equation without performing the secular approximation. This, however, can leave us vulnerable to the issue of violating the positivity of the density matrix in some cases as mentioned in the introduction. We do our best to work in parameter regimes with positive density matrices and comment if an issue arises. Quantification of the validity regime of the Bloch-Redfield equation will be investigated in future work.

\section{A competition perspective on steady-state entanglement}
\label{ent}

In the previous section, we have derived the quantum master equation beyond the secular approximation. With the dynamical equations for the density matrix elements given in the Appendix, the steady state can be obtained by solving the equation $d\rho/dt=0$. The specific expressions of the analytical solution of the steady state will be given later. In this section we only need the generic form of the steady-state density matrix.

Without the secular approximation, the diagonal elements (populations) of the density matrix $\rho_{ii}$ ($i=1, 2, 3, 4$) are in general coupled with a pair of off-diagonal elements (coherence) $\rho_{34} $ and $\rho_{43}$. As a consequence, at the steady state the density matrix in the energy eigen-state representation has the generic form
\begin{equation}
\rho_{ss}^{\rm{eig}}=
\left(\begin{array}{cccc}
\rho_{11} & 0 & 0 & 0\\
0 & \rho_{22} & 0 & 0\\
0 & 0 & \rho_{33} & \rho_{34}\\
0 & 0 & \rho_{43} & \rho_{44}
\end{array}\right).
\end{equation}
Here, $\rho_{34}$ and $\rho_{43}$ are the steady-state coherence induced by the nonequilibrium condition, which vanish at equilibrium as a consequence of the decoherence process.

Transformed into the bare-state representation with the basis $\{|ee\rangle,|gg\rangle,|eg\rangle,|ge\rangle\}$, the density matrix becomes
\begin{equation}
\rho_{ss}^{\rm{bar}}=U\rho_{ss}^{\rm{eig}}U^{\dagger}=
\left(\begin{array}{cccc}
a & 0 & 0 & 0\\
0 & d & 0& 0\\
0 & 0 & b & w\\
0 & 0 & w^{*} & c
\end{array}\right),\label{xstate}
\end{equation}
 where
\begin{subequations}
\begin{eqnarray}
a&=&\rho_{11},\, d=\rho_{22},\\
b&=&\cos^{2}\frac{\theta}{2}\rho_{33}+\sin^{2}\frac{\theta}{2}\rho_{44}-\frac{1}{2}\sin\theta
(\rho_{34}+\rho_{43}),\\
c&=&\sin^{2}\frac{\theta}{2}\rho_{33}+\cos^{2}\frac{\theta}{2}\rho_{44}+\frac{1}{2}\sin\theta
(\rho_{34}+\rho_{43}),\\
w&=&\frac{1}{2}\sin\theta(\rho_{33}-\rho_{44})+\cos^{2}\frac{\theta}{2}\rho_{34}-\sin^{2}
\frac{\theta}{2}\rho_{43}.\label{cohbsr}
\end{eqnarray}\label{bloch}
\end{subequations}
Here $w$ represents the quantum coherence in the bare-state representation. At equilibrium we have $\rho_{34}=\rho_{43}=0$ and thus $w=\sin\theta(\rho_{33}-\rho_{44})/2$, which in general does not vanish unless $\lambda=0$. In other words, due to the inter-qubit coupling, there is a residual quantum coherence in the bare-state representation even at equilibrium. When considering steady-state coherence induced by the nonequilibrium condition, we are always referring to the coherence in the eigen-state representation ($\rho_{34}$ and $\rho_{43}$)~\cite{coh2}.

The entanglement between the two qubits can be quantified by the concurrence~\cite{WK}. For the density matrix in the bare basis in Eq.~(\ref{xstate}), which is an example of the so-called ``X-state" (when the base vectors are arranged in the order $\{|ee\rangle,|eg\rangle,|ge\rangle,|gg\rangle\}$), the concurrence has the expression~\cite{MI,TY}
\begin{equation}
\mathcal{C}=2\max(0,|w|-\sqrt{ad}).
\label{concurrence}
\end{equation}
The state is entangled whenever $\mathcal{C}>0$ and maximally
entangled when $\mathcal{C}=1$. The expression of $\mathcal{C}$ suggests that the degree of entanglement, i.e. the value of concurrence, is determined by the competition between the coherence $w$ and the populations $a$ and $d$ in the bare-state representation. The coherence contributes positively to concurrence (the term $|w|$) and the populations contribute negatively to concurrence (the term $-\sqrt{ad}$). Only when the coherence dominates the populations (namely, $|w|>\sqrt{ad}$) will the state become entangled.

The underlying physics behind this literal interpretation of the concurrence expression can be better understood by writing the density matrix in Eq.~(\ref{xstate}) as $\rho=\widetilde{\rho}_1\oplus\widetilde{\rho}_2$, where
\begin{equation}
\widetilde{\rho}_1=\left(\begin{array}{cc}
a & 0\\
0 & d
\end{array}\right),\,\,
\widetilde{\rho}_2=\left(\begin{array}{cc}
b & w\\
w^{*} & c
\end{array}\right).
\end{equation}
Here $\widetilde{\rho}_1$ and $\widetilde{\rho}_2$ are ``density matrices'' (not normalized) in the Hilbert subspaces $\mathcal{H}_1$ and $\mathcal{H}_2$ spanned by $\{|ee\rangle,|gg\rangle\}$ and $\{|eg\rangle,|ge\rangle\}$, respectively. First consider the special case $\widetilde{\rho}_1=0$, i.e., $a=d=0$. The density matrix then reduces to $\widetilde{\rho}_2$ in $\mathcal{H}_2$ spanned by $\{|eg\rangle,|ge\rangle\}$. In this case, the expression of concurrence becomes $\mathcal{C}=2|w|$, which means the entanglement is essentially quantified by the coherence between $|eg\rangle$ and $|ge\rangle$ in $\mathcal{H}_2$. This is intuitively understandable given how the Bell states (maximally entangled states) are constructed as the coherent superpositions of the pair of separable states $|eg\rangle$ and $|ge\rangle$. When there are non-zero populations in $\mathcal{H}_1$ (i.e., $ad\neq 0$), the state of the system becomes more mixed since $\rho=\widetilde{\rho}_1\oplus\widetilde{\rho}_2$, or equivalently, $\rho=p_1\rho_1+p_2\rho_2$, where $\rho_1$ and $\rho_2$ ($4 \times 4$ matrices) are normalized versions of $\widetilde{\rho}_1$ and $\widetilde{\rho}_2$, respectively. It can be expected that the entanglement as a quantum property becomes weaker (at least not stronger) when the state becomes more mixed in a classical way. This is mathematically captured by the convexity property of concurrence, $\mathcal{C}(\rho)\leq p_1\mathcal{C}(\rho_1)+p_2\mathcal{C}(\rho_2)$, leading to $\mathcal{C}(\rho)\leq 2|w|$ in this case. (Note that we have used $\mathcal{C}(\rho_1)=0$, which can also be intuitively understood from the fact that $\rho_1$ is a mixture of the two separable states $|gg\rangle$ and $|ee\rangle$ without entanglement.) Therefore, the negative contribution of the populations in $\mathcal{H}_1$ to concurrence can be identified as the effect of mixing with separable states $|gg\rangle$ and $|ee\rangle$. This is the mechanism of competition between the coherence in $\mathcal{H}_2$ and the populations in $\mathcal{H}_1$ that determines the entanglement of the steady state of the system.

This competition perspective provides a basic structure to understand the behavior of the steady-state entanglement in both equilibrium and nonequilibrium settings as demonstrated in the following sections.

\section{entanglement and coherence in the equilibrium situation}
\label{equili}

In the equilibrium situation, the two baths share the same temperature and chemical potential. We consider the symmetric qubit case $\omega_1=\omega_2=\omega$ ($\theta=\pi/2$) here. (The asymmetric qubit case will be discussed in Sec.~\ref{phased}.) Our focus in this section for the equilibrium setting is on the physical understanding of the behaviors of entanglement and coherence less explored in the previous work.

It can be shown that in the eigen-state representation, the coherence at the equilibrium steady state vanishes, i.e., $\rho_{34}=\rho_{43}=0$, in agreement with decoherence. The populations can be found to be~\cite{en9}
\begin{eqnarray}
\rho_{11}&=&	\frac{(\gamma_{1}^{-}+\gamma_{2}^{-})(\gamma_{1}^{+}+\gamma_{2}^{+})}{Z},\label{rho11}\\
\rho_{22}&=&	\frac{(\Gamma_{1}^{-}+\Gamma_{2}^{-})(\Gamma_{1}^{+}+\Gamma_{2}^{+})}{Z},\\
\rho_{33}&=&\frac{(\Gamma_{1}^{-}+\Gamma_{2}^{-})(\gamma_{1}^{+}+\gamma_{2}^{+})}{Z}, \\
\rho_{44}&=&\frac{(\Gamma_{1}^{+}+\Gamma_{2}^{+})
(\gamma_{1}^{-}+\gamma_{2}^{-})}{Z},\label{rho44}
\end{eqnarray}
where
\begin{equation}
Z=(\gamma_{1}^{-}+\gamma_{2}^{-}+\Gamma_{1}^{-}+\Gamma_{2}^{-})
(\gamma_{1}^{+}+\gamma_{2}^{+}+\Gamma_{1}^{+}+\Gamma_{2}^{+})
\end{equation}
is the normalization factor. The above analytical solution for the equilibrium case can also be obtained as the limit of the analytical solutions we have obtained for the more general nonequilibrium scenarios given later.

According to Eq.~(\ref{bloch}), the coherence in the bare-state representation in this equilibrium setting reads
\begin{equation}\label{CohereEquil}
w=\frac{1}{2}(\rho_{33}-\rho_{44}).
\end{equation}
We will simply refer to $w$ as coherence \emph{in this section} given that the coherence in the eigen-state representation vanishes here. Notice that $|3\rangle=(|ge\rangle+|eg\rangle)/\sqrt{2}$ and $|4\rangle=(|ge\rangle-|eg\rangle)/\sqrt{2}$ (for $\theta=\pi/2$) are both maximally entangled and coherent states. The expression of $w$ in Eq.~(\ref{CohereEquil}) shows that these two states work against each other, in the sense that the coherence and thus entanglement (coherence $w$ contributes positively to entanglement) tend to decrease if the populations on these two states become more balanced and increase if more imbalanced. This may also be regarded as a ``competition perspective".

\subsection{Equilibrium Boson Reservoirs}

We first consider the case that the qubits are immersed in equilibrium boson reservoirs with $T_1=T_2=T$. According to Eq.~(\ref{BosonCoefficients}), we have
\begin{eqnarray}
\gamma_1^{+}=\gamma_2^{+}&=&\frac{J}{e^{\omega_{b+}/T}-1},\
\Gamma_1^{+}=\Gamma_2^{+}=\frac{Je^{\omega_{b+}/T}}{e^{\omega_{b+}/T}-1},\\
\gamma_1^{-}=\gamma_2^{-}&=&\frac{J}{e^{\omega_{b-}/T}-1},\
\Gamma_1^{-}=\Gamma_2^{-}=\frac{Je^{\omega_{b-}/T}}{e^{\omega_{b-}/T}-1},
\end{eqnarray}
with $\omega_{b\pm}=\omega\pm\lambda/2$ (in the symmetric qubit case $\delta=2\omega$, $\Omega=\lambda$ and $\lambda<2\omega$). The populations in the eigen-state representation can then be obtained from Eqs.~(\ref{rho11})-(\ref{rho44}), which agrees with the equilibrium canonical ensemble distribution $\rho_{ii}\propto e^{-E_i/T}$. Thus the coherence, according to Eq.~(\ref{CohereEquil}), has the explicit expression
\begin{equation}
w=-\frac{\sinh(\frac{\lambda}{2T})}
{2[\cosh(\frac{\omega}{T})+\cosh(\frac{\lambda}{2T})]},
\end{equation}
and the concurrence can then be obtained as
\begin{equation}\label{ConcurrenceEquilibrium}
\mathcal{C}=\max(0,\mathcal{E}_b)
\end{equation}
where
\begin{equation}\label{ConcurrenceEquilibriumEb}
\mathcal{E}_b=\frac{\sinh(\frac{\lambda}{2T})-1}
{\cosh(\frac{\omega}{T})+\cosh(\frac{\lambda}{2T})}.
\end{equation}

\begin{figure}[tbp]
\centering
\includegraphics[width=8cm]{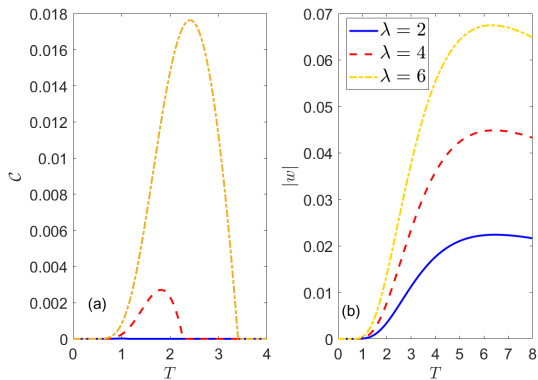}
\caption{(Color online) (a) The steady-state concurrence for the boson reservoir case. (b) The steady-state coherence in the bare-state representation for the boson reservoir case. The parameters are set as $\omega_1=\omega_2=10, J_1=J_2=1$.}
\label{bosoneq}
\end{figure}

The concurrence $\mathcal{C}$ and the coherence magnitude $|w|$ are plotted in Fig.~\ref{bosoneq}. As can be seen, concurrence and coherence both increase as the inter-qubit coupling strength becomes larger. This behavior can be expected as stronger inter-qubit coupling tends to enhance the quantum connection between the two qubits, making them more ``intertwined" with each other. It is also easy to show mathematically that $\mathcal{C}$ and $|w|$ monotonically increase with $\lambda$. Note that when $\lambda$ is small (e.g., the blue solid line in Fig.~{\ref{bosoneq}}(a) with $\lambda=2$), the concurrence graph appears to be completely flat, with zero value (i.e., no entanglement). However, the concurrence actually still has non-zero values in a temperature range, just too small to be seen in the figure. In fact, according to Eq.~(\ref{ConcurrenceEquilibriumEb}), the concurrence takes positive values when $\sinh(\lambda/2T)>1$, which leads to the temperature range $0<T<T_{\max}$ with $T_{\max}=\lambda/[2\ln(1+\sqrt{2})]$, no matter how small $\lambda$ is.

The concurrence and coherence (in the bare-state representation) are both non-monotonic functions of the temperature as can be seen in Fig.~{\ref{bosoneq}}(a) and (b). The non-monotonic behavior can be explained by considering the low and high temperature regimes respectively. In the low temperature regime, close to the absolute zero, the system is almost exclusively populated on the ground state $|2\rangle$=$|gg\rangle$ with zero entanglement and coherence. When the temperature is increased, the first excited state $|4\rangle=(|ge\rangle-|eg\rangle)/\sqrt{2}$ starts to become populated, while $|3\rangle=(|ge\rangle+|eg\rangle)/\sqrt{2}$ and $|1\rangle=|ee\rangle$ remain much less populated. Since $|3\rangle$ and $|4\rangle$ work against each other in the sense of Eq.~(\ref{CohereEquil}), the imbalance in their populations at this stage contributes to the increase of coherence. Moreover, since $|1\rangle$ is still underpopulated, the negative contribution of populations to concurrence ($\sqrt{ad}=\sqrt{\rho_{11}\rho_{22}}$) due to state mixing remains small. In other words, the coherence beats the populations in the battle of contributing to concurrence. These are the physical reasons why concurrence and coherence increase with temperature in the low temperature regime. Mathematically, for small $T$ we have
\begin{equation}
\mathcal{C}\approx 2|w|\approx\mathcal{E}_b\approx\frac{\sinh(\frac{\lambda}{2T})}
{\cosh(\frac{\omega}{T})}\approx e^{-\frac{\omega-\lambda/2}{T}},
\end{equation}
which increases with temperature (note $\omega>\lambda/2$).

In the high temperature regime, all the four eigen-states are fairly populated (equally populated in the limit $T\rightarrow \infty$). The more balanced populations on $|3\rangle$ and $|4\rangle$ weakens the coherence, resulting in the decreasing behavior of coherence with temperature in this regime. On the other hand, more balanced and significant populations on $|1\rangle$ and $|2\rangle$ enhance the negative contribution of populations to concurrence arising from state mixing. As a result, the populations take over in the competition with coherence, leading to zero concurrence in this regime. Mathematically, the Taylor expansion with respect to $1/T$ gives
\begin{equation}
\mathcal{E}_b\approx 2|w|-\frac{1}{2}\approx-\frac{1}{2}+\frac{\lambda}{4T}+o(1/T^2),
\end{equation}
which decreases with $T$ and becomes less than zero (so that $\mathcal{C}=0$) when $T$ is large enough. This approximate expression yields $T_{\max}\approx\lambda/2$, at which point the concurrence becomes zero. The exact temperature at the turning point was given previously, $T_{\max}=\lambda/[2\ln(1+\sqrt{2})]$. At this temperature the positive contribution by coherence and negative contribution by populations to concurrence are exactly balanced. For temperatures higher than $T_{\max}$, the populations in $|1\rangle$ and $2\rangle$ promoted by the thermal effect in the reservoirs beat the quantum coherence  in $|3\rangle$ and $|4\rangle$ between the two qubits, resulting in vanishing entanglement.

The maximum concurrence that can be achieved or asymptotically approached in this setting is of particular interest. For fixed $\lambda$, it is not easy to obtain an explicit analytical expression for the maximum of $\mathcal{C}(T)$. But we know that the concurrence increases monotonically with $\lambda$, and the upper bound of $\lambda$ is $2\omega$ due to the rotating wave approximation. (Actually the rotating wave approximation breaks down if $\lambda$ is close to $2\omega$). Thus we only need to consider the asymptotic case $\lambda\rightarrow 2\omega$, which leads to $\mathcal{E}_b=[\sinh(\omega/T)-1]/[2\cosh(\omega/T)]$, with its maximum value $1/2$ obtained as $T\rightarrow 0$. That means, in this equilibrium boson reservoir setting, the maximum concurrence that can be achieved cannot exceed $1/2$, one half of the maximum theoretical value of concurrence.

\subsection{Equilibrium Fermion Reservoirs}\label{EquFermBath}

To overcome the possible limitation in the approach of exchanging energy with the equilibrium boson reservoirs due to the thermal effects, we also investigate entanglement and coherence in relation to particle exchange (chemical potential) when the two qubits are immersed in equilibrium fermion reservoirs with $T_1=T_2=T$ and $\mu_1=\mu_2=\mu$. In this setting, we have
\begin{eqnarray}
\gamma_1^{+}=\gamma_2^{+}&=&\frac{J}{e^{\omega_{f+}/T}+1},
\Gamma_1^{+}=\Gamma_2^{+}=\frac{Je^{\omega_{f+}/T}}{e^{\omega_{f+}/T}+1},\\
\gamma_1^{-}=\gamma_2^{-}&=&\frac{J}{e^{\omega_{f-}/T}+1},
\Gamma_1^{-}=\Gamma_2^{-}=\frac{Je^{\omega_{f-}/T}}{e^{\omega_{f-}/T}+1},
\end{eqnarray}
with $\omega_{f\pm}=\omega-\mu\pm\lambda/2$. The coherence has the expression
\begin{equation}\label{CohEqFermi}
w=-\frac{\sinh(\frac{\lambda}{2T})}
{2[\cosh(\frac{\omega-\mu}{T})+\cosh(\frac{\lambda}{2T})]}
\end{equation}
and the concurrence is given by
\begin{equation}\label{ConEqFermi}
\mathcal{C}={\rm max}(0,\mathcal{E}_f)
\end{equation}
where
\begin{equation}
\mathcal{E}_f=\frac{\sinh(\frac{\lambda}{2T})-1}
{\cosh(\frac{\omega-\mu}{T})+\cosh(\frac{\lambda}{2T})}.\label{ef}
\end{equation}

\begin{figure}[tbp]
\centering
\includegraphics[width=8cm]{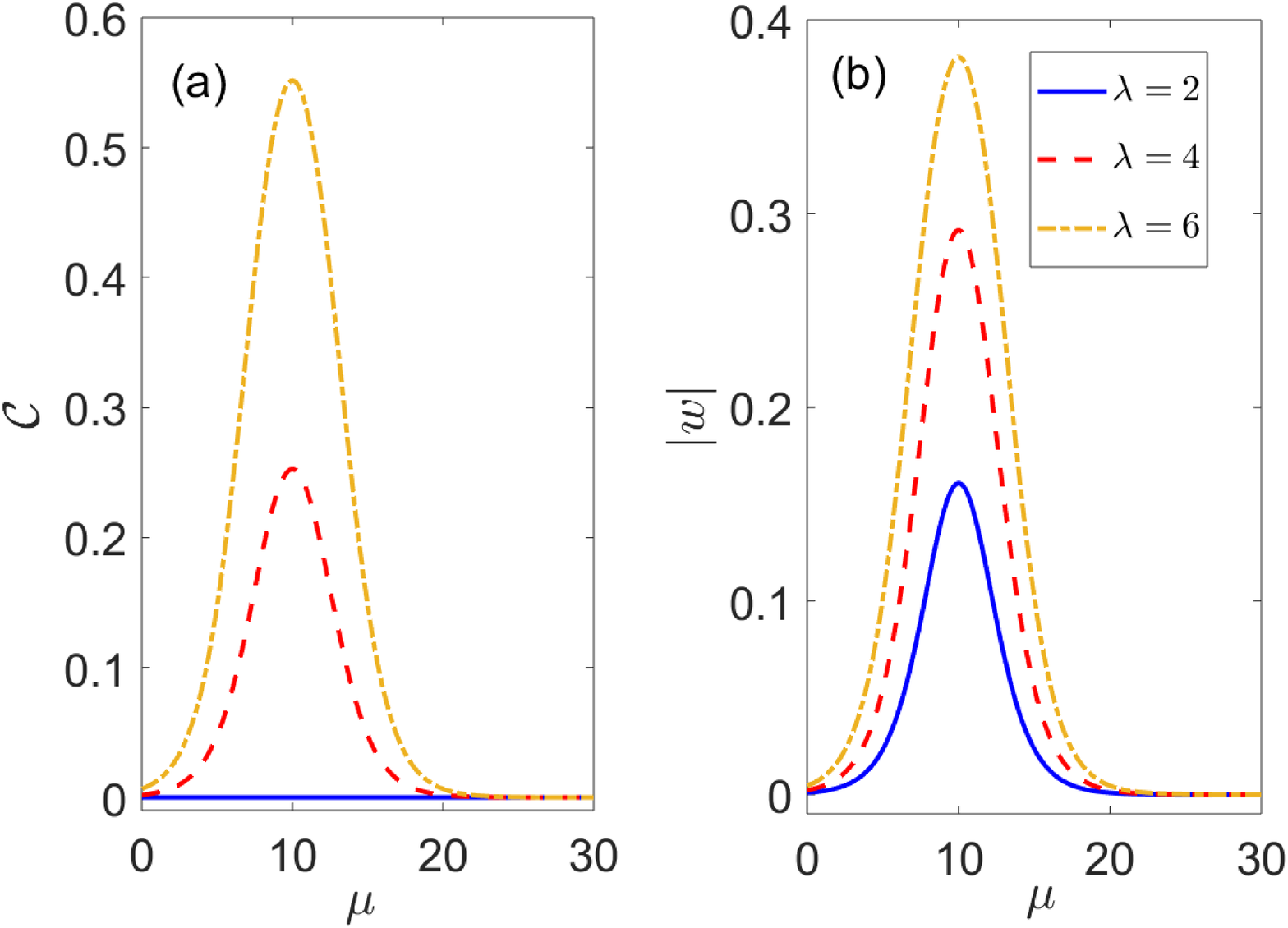}
\caption{(Color online) (a) The steady-state concurrence for the fermion reservoir case. (b) The steady-state coherence in the bare-state representation for the fermion reservoir case. The parameters are set as $\omega_1=\omega_2=10, J_1=J_2=1, T_1=T_2=1.5$.}
\label{fermioneq}
\end{figure}

The graphs of $\mathcal{C}$ and $|w|$ as functions of the chemical potential are shown in Fig.~\ref{fermioneq}, with the temperature fixed at a relatively low value $T=1.5$. As can be expected, both concurrence and coherence increase with the inter-qubit coupling strength in general. However, when $\lambda$ is too small (e.g., $\lambda=2$), the concurrence vanishes completely as can be seen in Fig.~(\ref{fermioneq})(a) (solid blue line). The concurrence here is genuinely zero in the entire range of $\mu$, in contrast with the solid blue line in Fig.~\ref{bosoneq}(a) where the concurrence only appears to be zero. According to Eq.~(\ref{ef}), the concurrence vanishes completely when $0<\lambda\leq\lambda_{\min}$ with $\lambda_{\min}=[2\ln(1+\sqrt{2})]T$. It is easy to check that $\lambda=2$ is in this range for $T=1.5$. In this range, the thermal effect due to the temperature of the reservoirs overrides the quantum connection established by the inter-qubit coupling, leading to vanishing entanglement.

Then we also notice in Fig.~(\ref{fermioneq}) that the concurrence and coherence (in the bare-state representation) are both non-monotonic functions of the reservoir chemical potential (when $\lambda>\lambda_{\min}$ for concurrence). The analytical expressions in Eqs.~(\ref{CohEqFermi}) and (\ref{ef}) show that the dependence on the chemical potential comes from the term $\cosh((\omega-\mu)/T)$ in the denominator, which has a minimum value at $\mu=\omega$ and reverses its monotonicity as $\mu$ crosses $\omega$. This is the mathematical reason why concurrence and coherence vary  non-monotonically with the chemical potential and have a maximum at $\mu=\omega$.

To gain a physical perspective on the features of concurrence and coherence in Fig.~(\ref{fermioneq}), we notice that in this equilibrium scenario the populations actually follow the grand canonical ensemble distribution $\rho_{ii}\propto e^{-(E_i-\mu N_i)/T}$, given that the system can change both energy and particle with the equilibrium reservoir. Here $N_i$ is the particle number on the energy level $E_i$. What is relevant is the particle number difference between energy levels instead of their absolute values. The form of the qubit-reservoir interaction Hamiltonian in Eq.~(\ref{Hamiltonian}) indicates that the particle number in the excited state $|e\rangle$ of a qubit is one larger than that in the ground state $|g\rangle$ due to particle absorption from the reservoir. This translates into the particle number assignment $N_1=2$, $N_2=0$, $N_3=N_4=1$ for the eigen-states. Up to a normalization factor, the grand canonical distribution then yields $\rho_{11}\propto e^{-(\omega-\mu)/T}$, $\rho_{22}\propto e^{(\omega-\mu)/T}$, $\rho_{33}\propto e^{-\lambda/2T}$, $\rho_{44}\propto e^{\lambda/2T}$. This suggests the introduction of the `effective' energies $\pm(\omega-\mu)$ and $\pm \lambda/2$ associated with $|1\rangle$, $|2\rangle$, $|3\rangle$, $|4\rangle$, respectively, which determine the populations on these eigen-states. The chemical potential simply adjusts the effective energy level spacing between $|1\rangle$ and $|2\rangle$ while $|3\rangle$ and $|4\rangle$ are fixed. As $\mu$ increases, $|1\rangle$ and $|2\rangle$ become effectively closer, and coincide with each other at $\mu=\omega$, after which they split up again. This process is symmetric with respect to the cross-over point $\mu=\omega$, as far as concurrence and coherence are concerned. This explains the feature that $\mathcal{C}$ and $|w|$ are symmetric with respect to $\mu=\omega$ (with $\omega=10$) in Fig.~\ref{fermioneq} (for fermion reservoirs $\mu$ can also take negative values). Moreover, as $|1\rangle$ and $|2\rangle$ get closer to each other with increasing $\mu$, the combined weights of the populations on $|1\rangle$ and $|2\rangle$ decrease. (This can be understood by considering the extreme cases of a very large and a very small level spacing.) Roughly speaking, this means there is a population transfer from $|1\rangle$ and $|2\rangle$ together to $|3\rangle$ and $|4\rangle$. In turn it suggests an increase in entanglement and coherence, given that $|3\rangle$ and $|4\rangle$ are entangled and coherent states while $|1\rangle$ and $|2\rangle$ are separable and incoherent states. This provides an intuitive perspective on the increasing interval of entanglement and coherence as $\mu<\omega$. When $\mu$ further increases beyond $ \omega$, the process is reversed according to the symmetry argument above. That is, the separable and incoherent states $|1\rangle$ and $|2\rangle$ together become more populated, while the entangled and coherent states $|3\rangle$ and $|4\rangle$ become less populated. This suggests the decrease of entanglement and coherence with $\mu$ as $\mu>\omega$. Thus we can intuitively understand the non-monotonic behaviors of entanglement and coherence in Fig. 4.

The maximum concurrence is achieved at $\mu=\omega$ with the value $\mathcal{C}_{\max}=[\sinh(\lambda/2T)-1]/[1+\cosh(\lambda/2T)]$ for $\lambda>\lambda_{\min}$. As the temperature $T$ approaches the absolute zero, this value approaches one, the theoretical maximum value of concurrence. In other words, in this equilibrium fermion reservoir setting, maximally entangled state with concurrence one can be approached as $T\rightarrow 0$ with $\mu=\omega$ and fixed positive $\lambda$. The physics is easy to understand in the picture of the effective energy introduced in the previous paragraph. At $\mu=\omega$, the effective energies of $|1\rangle$ and $|2\rangle$ both become zero, which means these two levels effectively coincide with each other in between $|3\rangle$ and $|4\rangle$ (with effective energies $\pm\lambda/2$). Thus the effective ground state with the lowest effective energy becomes $|4\rangle=(|ge\rangle-|eg\rangle)/\sqrt{2}$, a maximally entangled state with concurrence one. At absolute zero, the system is only populated in $|4\rangle$, thus maximally entangled.

Judging from the maximum concurrence that can be asymptotically achieved or approached, equilibrium fermion reservoir with particle exchange (maximum concurrence $1$) is more beneficial to enhancing quantum entanglement than equilibrium boson reservoir with energy exchange (maximum concurrence not exceeding $1/2$). However, since these two settings use different approaches of system-reservoir exchange (energy exchange versus particle exchange), one may wonder to what extent the difference between the statistics in these two settings played a role. To place the two types of reservoirs on an equal footing, we may also consider boson reservoirs with a chemical potential that can exchange particle with the system. At equilibrium, the expressions of the populations and thus the coherence and concurrence in this boson bath setting are actually the same as those in the fermion setting in Eqs.~(\ref{CohEqFermi})-(\ref{ef}), in agreement with the grand canonical ensemble. However, there is one important difference. That is, the chemical potential of the boson reservoir is negative (due to the Boson-Einstein statistics), while there is no such constraint on the fermion reservoir chemical potential. This sign restriction on the boson reservoir chemical potential prevents it from taking the value $\mu=\omega$ to achieve maximum concurrence as in the fermion case. Therefore, the difference in statistics did play an important role in manifesting different entanglement behaviors in these two equilibrium settings.

\section{entanglement and coherence in the nonequilibrium situation}
\label{nonequili}

In the pervious section we investigated the steady-state entanglement and
coherence for the equilibrium situation. We now consider the nonequilibrium setup when the two reservoirs have different temperatures (boson reservoirs) or chemical potentials (fermion reservoirs). We still focus on the symmetric qubit case $\omega_1=\omega_2=\omega$ ($\theta=\pi/2$, $\delta=2\omega$, $\Omega=\lambda$). We have obtained the analytical expressions of the steady-state density matrix for the nonequilibrium scenario, using the method outlined in Appendix \ref{appendix2}. To express the analytical solutions in a concise form, we introduce the following notations
\begin{eqnarray}
&&\bar{N}_+=\frac{1}{2}(N_1^++N_2^+),\quad \bar{N}_-=\frac{1}{2}(N_1^-+N_2^-),\label{NBar}\\
&&\widetilde{N}_+=\frac{1}{2}(N_1^+-N_2^+),\quad \widetilde{N}_-=\frac{1}{2}(N_1^--N_2^-),\label{NTilde}
\end{eqnarray}
where $N_i^{\pm}=N_i(\delta/2\pm \Omega/2)$ and $N_i(\omega)$ is the average particle number on frequency $\omega$ in the $i$-th reservoir that obeys Bose-Einstein or Fermi-Dirac statistics. Notice that $\widetilde{N}_{\pm}=0$ at equilibrium and in general $\widetilde{N}_{\pm}\neq 0$ in a nonequilibrium setting when the two reservoirs have different temperatures or chemical potentials. This means $\widetilde{N}_{\pm}$ can be considered as indicators of the nonequilibrium condition. On the other hand, $\bar{N}_{\pm}$ as the particle number averaged between the two baths represent a form of average equilibrium effect of the two baths.

\subsection{Nonequilibrium Boson Reservoirs}

We consider the two coupled qubits embedded in their individual boson reservoirs at different temperatures. The nonequilibrium condition is characterized by the temperature difference $\Delta T=T_2-T_1$.

The expressions of the steady-state populations in the eigen-state representation are obtained as
\begin{eqnarray}
\rho_{11}&=&\frac{1}{\mathcal{N}}\left[\bar{N}_+\bar{N}_--r_1r_2R\right],\label{PopulationBB11}\\
\rho_{22}&=&\frac{1}{\mathcal{N}}\left[(1+\bar{N}_+)(1+\bar{N}_-)-s_1s_2R\right],\\
\rho_{33}&=&\frac{1}{\mathcal{N}}\left[\bar{N}_+(1+\bar{N}_-)+s_1r_2R\right],\\
\rho_{44}&=&\frac{1}{\mathcal{N}}\left[\bar{N}_-(1+\bar{N}_+)+s_2r_1R\right],\label{PopulationBB44}
\end{eqnarray}
and the steady-state coherence in the eigen-state representation has the expression
\begin{equation}\label{NBRCER}
\rho_{34}=-\frac{1}{\mathcal{N}}\left[\frac{\widetilde{N}_+(1+2\bar{N}_-)+\widetilde{N}_-(1+2\bar{N}_+)}{2(1+\bar{N}_++\bar{N}_-) +i\Omega'}\right],
\end{equation}
where
\begin{eqnarray}
r_1&=&\widetilde{N}_++\widetilde{N}_-(1+2\bar{N}_++2\bar{N}_-),\\
r_2&=&\widetilde{N}_-+\widetilde{N}_+(1+2\bar{N}_++2\bar{N}_-),\\
s_1&=&\widetilde{N}_+-\widetilde{N}_-(3+2\bar{N}_++2\bar{N}_-),\\
s_2&=&\widetilde{N}_--\widetilde{N}_+(3+2\bar{N}_++2\bar{N}_-),\\
R&=&\frac{1}{4(1+\bar{N}_++\bar{N}_-)^2+\Omega'^2}.
\end{eqnarray}
In the above, $\Omega'=\Omega/J$ is the rescaled Rabi frequency and $\mathcal{N}$ is the normalization factor given by
\begin{equation}\label{NormBB}
\mathcal{N}=(1+2\bar{N}_+)(1+2\bar{N}_-)-16\widetilde{N}_+\widetilde{N}_-(1+\bar{N}_++\bar{N}_-)^2R.
\end{equation}
Accordingly, the coherence in the bare-state representation (for the symmetric qubit case) is given by
\begin{equation}\label{coherenceSQ}
w=\frac{1}{2}(\rho_{33}-\rho_{44})+i\text{Im}\rho_{34},
\end{equation}
and the concurrence can be obtained from
\begin{equation}\label{concurrenceSQ}
\mathcal{C}=2\max\{0, |w|-\sqrt{ad}\}
\end{equation}
with $a=\rho_{11}$ and $d=\rho_{22}$.

The steady-state solution in Eqs.~(\ref{PopulationBB11})-(\ref{NBRCER}) has the structure of an average equilibrium solution plus nonequilibrium corrections. It is easy to check that this nonequilibrium solution reduces to the equilibrium one when $\Delta T=0$ by noticing that $\widetilde{N}_{\pm}=0$ so that $r_1=r_2=s_1=s_2=0$ and $\rho_{34}=0$ at equilibrium. In this sense, the terms involving $r_1$, $r_2$, $s_1$, $s_2$ and $\rho_{34}$ represent nonequilibrium corrections. The rest of the terms represent the average equilibrium solution determined by $\bar{N}_{\pm}$. For the boson bath case here (the chemical potential is zero), it may be suggestive to introduce an effective temperature $T_{\text{eff}}$ defined in terms of $\bar{N}(\omega)=[N_1(\omega)+N_2(\omega)]/2$ by the Bose-Einstein statistics $\bar{N}(\omega)=1/(e^{\omega/T_{\text{eff}}}-1)$. However, the effective temperature $T_{\text{eff}}$ defined this way, in general, is not only determined by $T_1$ and $T_2$ of the two baths, but also dependent on $\omega$.  This means $\bar{N}(\omega)$, in general, cannot be truly associated with an effective equilibrium bath with a constant temperature $T_{\text{eff}}$. But in the near equilibrium regime ($\Delta T=T_2-T_1$ is small) or in the high temperature regime ($T_1$ and $T_2$ are both high), we have
\begin{equation}\label{ET}
T_{\text{eff}}\approx \bar{T}=(T_1+T_2)/2,
\end{equation}
which is a constant temperature independent of $\omega$. Therefore, in these regimes the average equilibrium solution may be interpreted as effectively generated by an equilibrium bath with the average temperature of the two baths.

\begin{figure}[tbp]
\centering
\includegraphics[width=8cm]{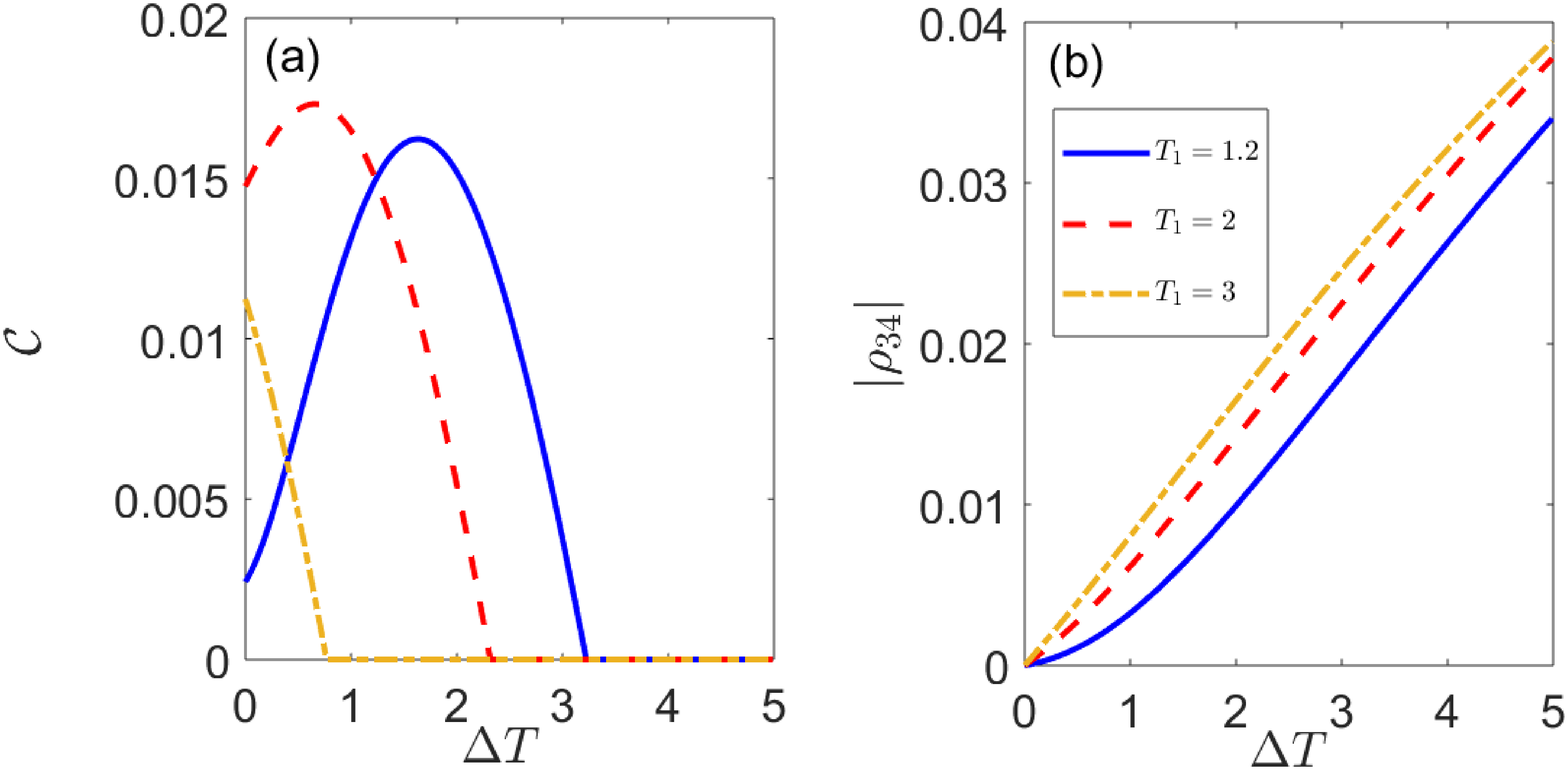}
\includegraphics[width=8cm]{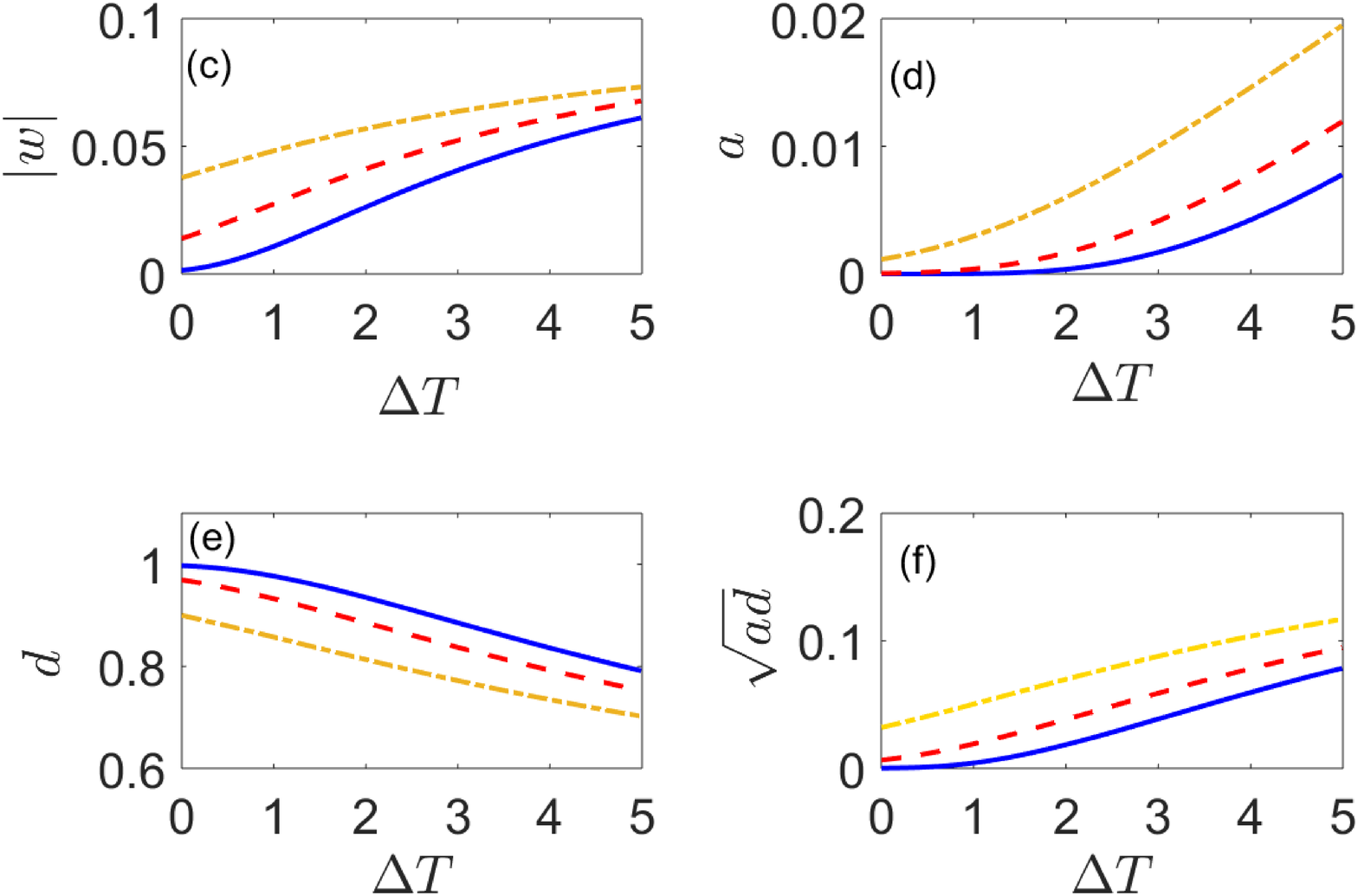}
\caption{(Color online) The steady-state concurrence, coherence in the eigen-state and bare-state representations, populations $a=\rho_{11}$ and $d=\rho_{22}$ as well as $\sqrt{ad}$, for the coupled qubit system interacting with individual boson reservoirs at different temperatures. The parameters are set as $\omega_1=\omega_2=10, J_1=J_2=1, \lambda=6$.}
\label{bosonneq}
\end{figure}

Our numerical calculations produce consistent results with the analytical solution. The concurrence and the coherence in the eigen-state representation are plotted in Fig.~\ref{bosonneq} (a) and (b), respectively. To understand the behavior of the concurrence in terms of the competition between $|w|$ and $\sqrt{ad}$, we also plotted $|w|$, $a$, $d$ and $\sqrt{ad}$ in Fig.~\ref{bosonneq} (c), (d), (e) and (f), respectively. The temperature difference is restricted to a relatively small range $\Delta T \in [0, 5]$ to conform to physically reasonable temperature conditions in quantum physics experiments and avoid possible artifacts that may arise from applying the Bloch-Redfield equation outside its valid parameter ranges.

As can be seen in Fig.~\ref{bosonneq} (b), the coherence in the eigen-state representation $\rho_{34}$, which represents the coherence induced by the nonequilibrium condition, increases monotonically with the nonequilibrium condition characterized by the temperature difference $\Delta T$. (This monotonic behavior continues even when $\Delta T$ becomes large.) The analytical expression of $\rho_{34}$ in Eq.~(\ref{NBRCER}) dictates its behavior with respect to $\Delta T$ given its dependence on the nonequilibrium indicators $\widetilde{N}_{\pm}$, which vanish at equilibrium and start to increase in magnitude when displaced from equilibrium. In Fig.~\ref{bosonneq} (b) we can also see that for fixed $\Delta T$, the value of $|\rho_{34}|$ increases with the base temperature $T_1$. However, as $T_1$ becomes higher (not shown in the figure), this behavior becomes reversed (i.e., $|\rho_{34}|$ will decrease with $T_1$) and $|\rho_{34}|$ becomes very small when $T_1$ is very high. This can be understood as the thermal effect prevailing over the quantum effect (coherence) when the temperatures of both baths are high.

The concurrence, as shown in Fig.~\ref{bosonneq} (a), in general displays non-monotonic behaviors with respect to the temperature difference. More specifically, when the base temperature $T_1$ is relatively low (e.g., $T_1=1.2$ and $T_1=2$ in the figure), the concurrence first increases and then decreases with $\Delta T$ until it hits zero. When $T_1$ becomes higher (e.g., $T_1=3$), however, the concurrence decreases monotonically with $\Delta T$. Moreover, when $T_1$ is further increased (e.g., $T_1=5$, not shown in the figure), the concurrence vanishes completely, irrespective of the temperature difference $\Delta T$.

The behavior of concurrence can be understood from the perspective of the competition between $|w|$ (the coherence between $|eg\rangle$ and $|ge\rangle$) and $\sqrt{ad}$ (the populations on $|ee\rangle$ and $|gg\rangle$). $|w|$ and $\sqrt{ad}$ both increase monotonically with $\Delta T$ as shown in Fig.~\ref{bosonneq} (c) and (f), respectively. However, how fast they increase with $\Delta T$ is dependent on the base temperature $T_1$ as  well as the value of $\Delta T$.

For a low base temperature, when $\Delta T$ is small, which implies that the temperatures of both baths are low, the system is mainly populated on the ground state, with $d\approx 1$ and $a\approx 0$ resulting in $\sqrt{ad}\approx0$, as can also be seen in Fig.~\ref{bosonneq} (d), (e) and (f) for $T_1=1.2$. In this regime, the concurrence is dominated by the behavior of the coherence $|w|$ that increases with $\Delta T$, leading to the monotonically increasing interval of the concurrence. When $\Delta T$ becomes larger, however, $\sqrt{ad}$ increases faster with  $\Delta T$ than $|w|$ does, resulting in the monotonically decreasing interval of the concurrence until it vanishes as $\sqrt{ad}\geq |w|$.

For a higher base temperature (e.g., $T_1=3$), even in the small $\Delta T$ regime, the temperatures of both baths ($T_2\geq T_1$) have already allowed the system to populate less unevenly between the ground state and the excited state, so that $\sqrt{ad}$ is not that small as in the case $T_1=1.2$ for the behavior of $|w|$ to dominate. Instead, $\sqrt{ad}$ increases with $\Delta T$ faster than $|w|$ in the range of $\Delta T$ considered and thus the concurrence decreases monotonically with $\Delta T$ in this regime.

For an even higher base temperature (e.g., $T_1=5$), the ground state and the excited states will be more evenly populated (which becomes almost uniformly distributed in extremely high temperatures).  As a result, the population term $\sqrt{ad}$ overrides the coherence term $|w|$ for all $\Delta T$, leading to vanishing concurrence irrespective of $\Delta T$. This is the thermal effect beating the quantum connection (entanglement) in high temperatures. This result also suggests that, to exploit the nonequilibrium condition (in this case the temperature difference) in enhancing entanglement, the temperature of at least one bath needs to be low enough to make sure that the thermal effect does not dominate; otherwise it will be futile to merely adjust the nonequilibrium condition.

Another perspective to understand the behavior of the concurrence in the near equilibrium regime (i.e., when $\Delta T$ is very small) is based on the effective temperature in Eq.~(\ref{ET}). In the near equilibrium regime, the nonequilibrium corrections are very small, so that the average equilibrium solution dominates, which means the system behaves almost like an equilibrium one, with the equilibrium temperature replaced by the average temperature $\bar{T}=(T_1+T_2)/2$ of the two baths. With $T_1$ fixed, increasing $T_2$ (thus increasing $\Delta T$) means the average temperature $\bar{T}$ starts to increase from $T_1$. We have already discussed how the concurrence behaves with temperature in the equilibrium boson bath case.  As can be seen in Fig.~\ref{bosoneq} (a) (the line for $\lambda=6$), the concurrence will start to either increase or decrease with the temperature, depending on its initial value. In this particular case ($\lambda=6$), the temperature at which the concurrence changes behavior from increasing to decreasing is around $T\approx 2.4$. This means for $T_1<2.4$, the concurrence will start to increase as $\Delta T$ is increased, while for $T_1>2.4$ it will start to decrease. This explains why in Fig.~\ref{bosonneq} (a) the lines for $T_1=1.2$ and $T_1=2$ have an initially increasing segment, while that for $T_1=3$ does not.

In addition, we remark that our analytical solution and numerical calculation indicate that when the base temperature is not too low, as for the cases considered in Fig.~\ref{bosonneq}, the concurrence will revive when the temperature difference $\Delta T$ becomes very large and it approaches non-vanishing values in the limit $\Delta T\rightarrow \infty$. (The analytical solution in this case can be obtained by replacing $N_2^{\pm}$ with its asymptotic form at high temperature, $T_2/(\omega\pm\lambda/2)$, and then taking the limit $T_2\rightarrow \infty$.) However, we cannot be certain whether this is a genuine physical effect or merely  an artifact created by applying the Bloch-Redfield equation out of its range of validity. A further investigation on this issue with solutions to the exact dynamics is certainly worthwhile, which is, however, beyond the scope of the present paper.

\subsection{Nonequilibrium Fermion Reservoirs}

Then we consider the two coupled qubits in contact with their individual fermion reservoirs with the same temperature $T_1=T_2$ but different chemical potentials $\mu_1\neq \mu_2$. The nonequilibrium condition is characterized by the chemical potential difference $\Delta \mu=\mu_2-\mu_1$.

The analytical solution of the steady-state density matrix for the fermion bath case is even simpler than that for the boson bath case. The steady-state populations in the eigen-state representation have the expressions
\begin{eqnarray}
\rho_{11}&=&\bar{N}_+\bar{N}_--\widetilde{R},\label{PopulationFBG11}\\
\rho_{22}&=&(1-\bar{N}_+)(1-\bar{N}_-)-\widetilde{R},\\
\rho_{33}&=&\bar{N}_+(1-\bar{N}_-)+\widetilde{R},\\
\rho_{44}&=&\bar{N}_-(1-\bar{N}_+)+\widetilde{R},\label{PopulationFBG44}
\end{eqnarray}
and the steady-state coherence in the eigen-state representation is given by
\begin{equation}\label{rho34FBG}
\rho_{34}=-\frac{\widetilde{N}_++\widetilde{N}_-}{2+i\Omega'},
\end{equation}
where
\begin{equation}\label{ExpR}
\widetilde{R}=|\rho_{34}|^2=\frac{(\widetilde{N}_++\widetilde{N}_-)^2}{4+\Omega'^2}
\end{equation}
and $\Omega'=\Omega/J$. Accordingly, the coherence in the bare-state representation $w$ and the concurrence $\mathcal{C}$ can be calculated using Eqs.~(\ref{coherenceSQ}) and (\ref{concurrenceSQ}), respectively.

The analytical solution also has the clear structure of an average equilibrium solution (associated with $\bar{N}_{\pm}$) corrected by nonequilibrium contributions (associated with $\widetilde{N}_{\pm}$). At equilibrium with $\mu_1=\mu_2$ and $T_1=T_2$, we have $\widetilde{N}_{\pm}=0$ and thus $\rho_{34}=\widetilde{R}=0$, reducing the nonequilibrium solution to the equilibrium one. $\widetilde{R}$ and $\rho_{34}$ represent nonequilibrium corrections, while the rest represent the average equilibrium solution. We may introduce an effective chemical potential $\mu_{\text{eff}}$ defined in terms of $\bar{N}(\omega)=[N_1(\omega)+N_2(\omega)]/2$ by the Fermi-Dirac statistics $\bar{N}(\omega)=1/(e^{(\omega-\mu_{\text{eff}})/T}+1)$ where $T=T_1=T_2$. In general, $\bar{\mu}$ is dependent on $\mu_1$ and $\mu_2$ as well as $\omega$ and $T$. In the near equilibrium regime ($\Delta\mu$ is small), we have
\begin{equation}
\mu_{\text{eff}}\approx \bar{\mu}=(\mu_1+\mu_2)/2.
\end{equation}
Then the average equilibrium solution in the near equilibrium regime may be interpreted as being generated by an effective equilibrium fermion reservoir with temperature $T$ and chemical potential $\bar{\mu}$.

\begin{figure}[tbp]
\centering
\includegraphics[width=8cm]{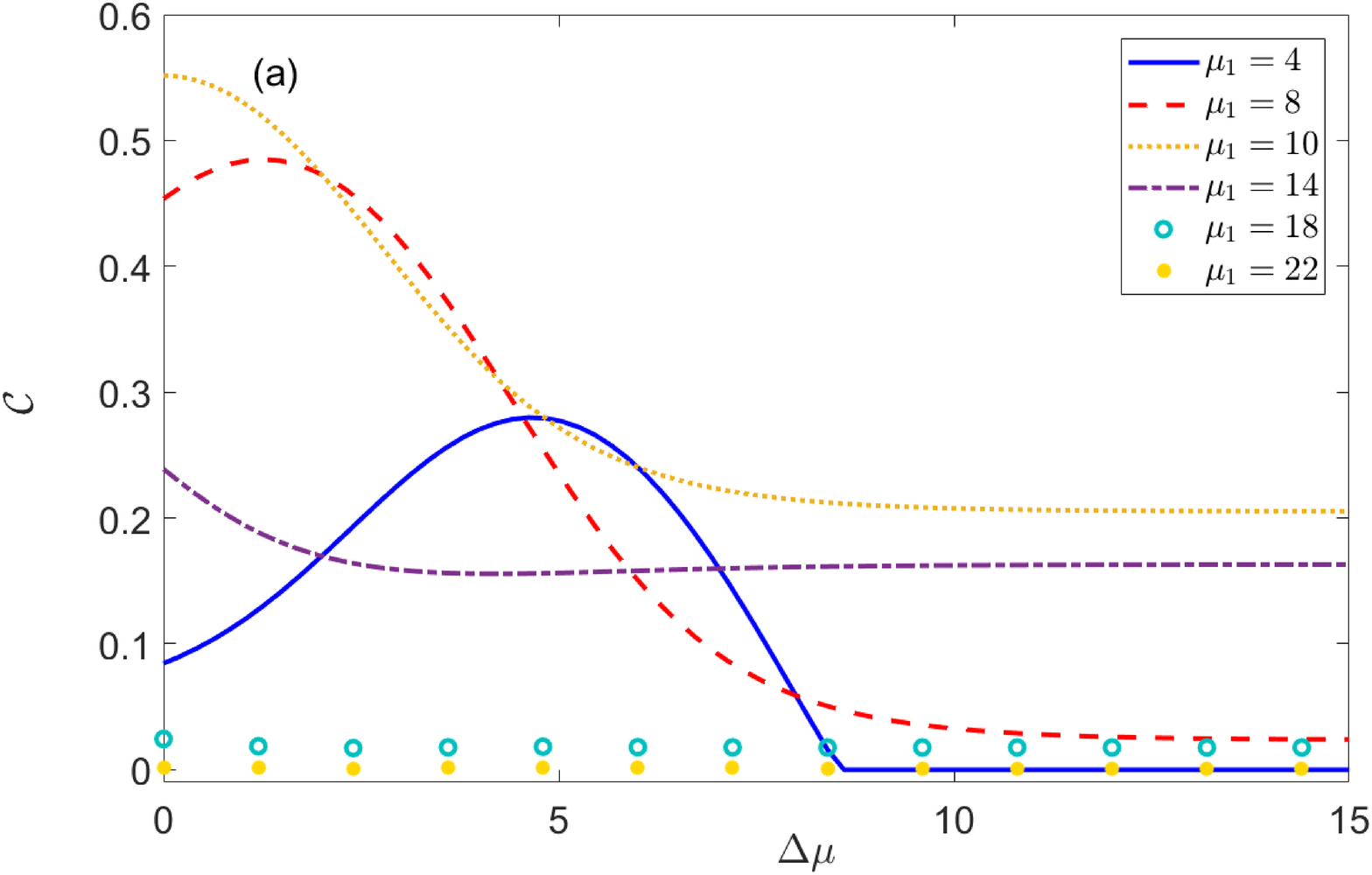}
\includegraphics[width=8cm]{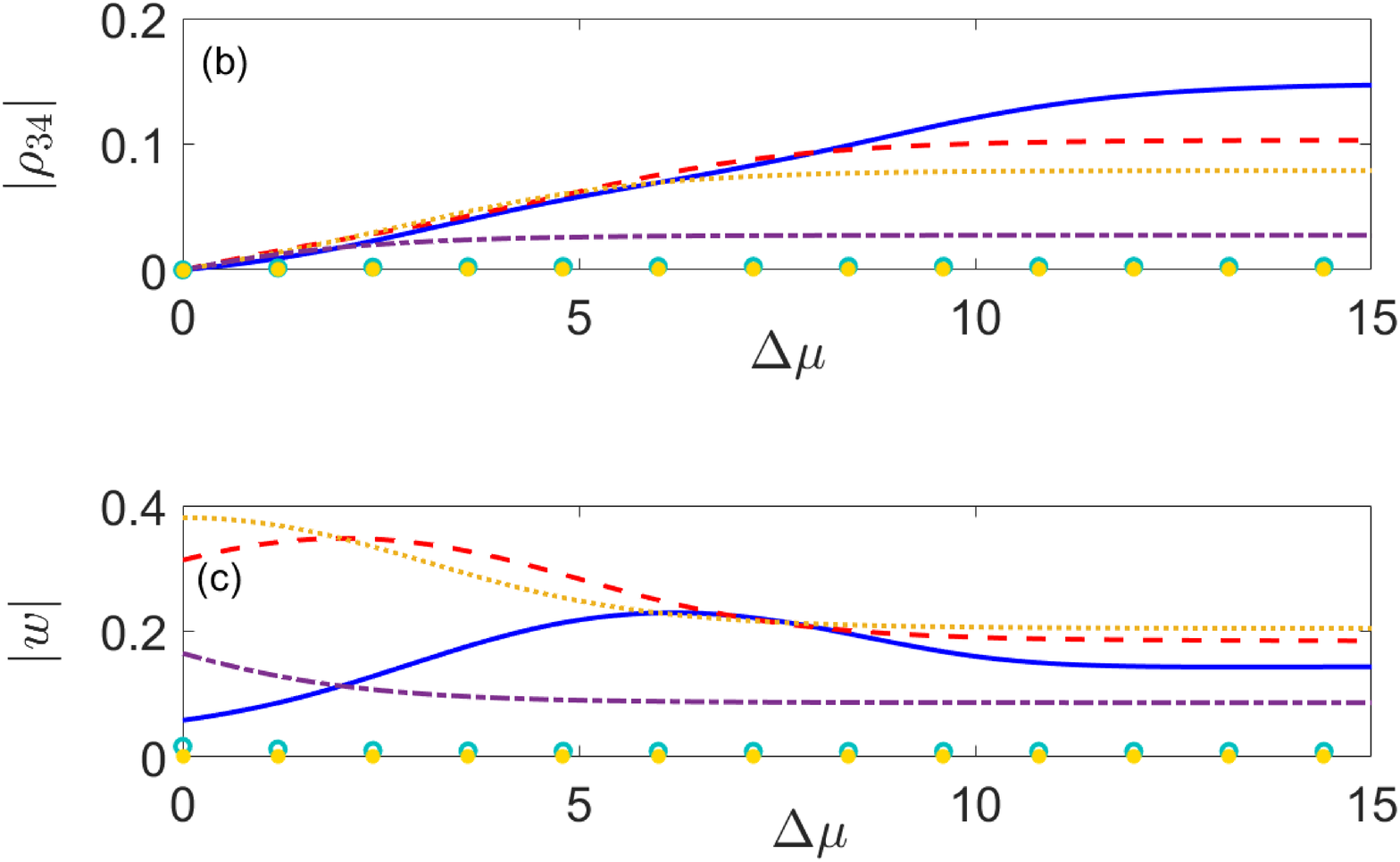}
\caption{(Color online) The steady-state concurrence (a), the coherence in the eigen-state representation (b), the coherence in the bare-state representation (c) for the coupled qubit system immersed in individual fermion reservoirs with different chemical potentials. The parameters are set as $\omega_1=\omega_2=10, J_1=J_2=1, \lambda=6, T_1=T_2=1.5$.}
\label{fermionneq}
\end{figure}

The concurrence and the coherence in the eigen-state and bare-state representations as functions of $\Delta \mu$  for different values of $\mu_1$ are plotted in Fig.~\ref{fermionneq} (a), (b) and (c), respectively. The parameters of $\omega$ and $\lambda$ are the same as those for the boson bath case. The temperatures of the two fermion baths, set equal to each other ($T_1=T_2=1.5$), have the same order of magnitude as those for the boson bath case. This is to make the fermion bath case more or less comparable with the boson bath case, although the nonequilibrium conditions in these two cases are different (temperature difference versus chemical potential difference).

We first have a look at the coherence $\rho_{34}$ induced by the nonequilibrium condition illustrated in Fig.~\ref{fermionneq} (b). As one can see, the magnitude of $\rho_{34}$ increases monotonically with the nonequilibrium condition characterized by $\Delta \mu$. When $\Delta \mu$ becomes large enough, $|\rho_{34}|$ approaches some asymptotic values depending on the value of $\mu_1$. Also, for a large base chemical potential $\mu_1$, the nonequilibrium-induced coherence $\rho_{34}$ becomes very small regardless of the nonequilibrium condition $\Delta \mu$.

On the other hand, the concurrence shown in Fig.~\ref{fermionneq} (a) has a more complicated behavior with respect to $\Delta \mu$, depending on the value of the base chemical potential $\mu_1$. For small $\mu_1$ ($\mu_1<\omega$), the concurrence is a non-monotonic function of $\Delta \mu$, which first increases and then decreases with $\Delta \mu$. For a larger $\mu_1$ ($\mu_1 >\omega)$, the concurrence decreases monotonically from its equilibrium value as $\Delta \mu$ increases. As $\mu_1$ becomes large enough (e.g. $\mu_1=22$), the concurrence is significantly suppressed no matter how large $\Delta \mu$ is, similar to the behavior of the coherence $\rho_{34}$. In addition, the concurrence also approaches some fixed values as $\Delta \mu$ grows large enough.

The common behavior of the coherence $\rho_{34}$ and the concurrence at large base chemical potential $\mu_1$ (both are significantly suppressed regardless of $\Delta \mu$) can be understood as follows. Note that $\Delta \mu\geq 0$, which means a large $\mu_1$ also implies a large $\mu_2$. That is, the chemical potential of both fermion baths are large. For large chemical potentials, we have $N_1^+\approx N_1^-\approx N_2^+\approx N_2^-\approx 1$, which implies that the nonequilibrium indicators $\widetilde{N}^{\pm}\approx 0$. Therefore, this situation is almost like an equilibrium one, regardless of the value of $\Delta \mu$. This explains the vanishing behavior of $\rho_{34}$ at large $\mu_1$. Moreover, given that particles tend to flow from higher chemical potentials to lower ones, the large chemical potentials in both baths mean the two qubits are both likely to be excited to the state $|e\rangle$ by the particle influx from the reservoirs, so that the system is mainly populated on the separable state $|ee\rangle$ (i.e., $\rho_{11}\approx 1$), leading to diminished entanglement. This accounts for the suppressed concurrence at large $\mu_1$. This result suggests that, in order to exploit the nonequilibrium condition (chemical potential difference) to effectively enhance steady-state entanglement, the chemical potential of one bath needs to be low enough.

As to the asymptotic behavior of the concurrence and the coherence for large $\Delta \mu$, this belongs to the strong nonequilibrium regime. Notice that $\mu_1$ is fixed and large $\Delta \mu$ simply means large $\mu_2$. The particular physical condition here is that the second bath becomes saturated due to its high chemical potential, so that particles mainly flow from the second bath to the system and rarely the other way around. Mathematically, as $\Delta\mu\rightarrow\infty$ ($\mu_2\rightarrow\infty$), we have $N_2^+=N_2^-=1$, yielding $\bar{N}_{\pm}=(N_1^{\pm}+1)/2$ and $\widetilde{N}_{\pm}=(N_1^{\pm}-1)/2$. Inserting them into the analytical solution, we obtain the asymptotic expressions of the steady-state solution for large $\Delta \mu$, which depends on the value of $\mu_1$ through $N_1^{\pm}$. The physical condition of particle saturation in the second bath and the asymptotic analytical solution can account for the features of coherence and entanglement at large $\Delta \mu$. However, as with the boson bath case, we are not certain whether the results in this regime are reliable.

Then we consider the opposite regime, namely, the near equilibrium regime when $\Delta \mu$ is very small. In particular, we want to understand the feature of the concurrence in this regime. (The coherence $\rho_{34}$ is expected to increase with $\Delta \mu$ in this regime as it represents nonequilibrium corrections.) In this regime the nonequilibrium corrections are not significant and the solution is dominated by the effective equilibrium part. This means the system behaves almost like an equilibrium system with a chemical potential $\bar{\mu}=(\mu_1+\mu_2)/2$. With $\mu_1$ fixed, increasing $\Delta\mu$ from zero is equivalent to increasing $\bar{\mu}$ from $\mu_1$. From the study in the equilibrium fermion bath case as illustrated in Fig.~\ref{fermioneq} (a), we know that the concurrence is a non-monotonic function of $\mu$ maximized at $\mu=\omega$, with $\mu<\omega$ the increasing interval and $\mu>\omega$ the decreasing interval. Therefore, when $\mu_1<\omega$, increasing $\Delta \mu$ will increase the concurrence, while the opposite is true when $\mu_1>\omega$, as long as $\Delta\mu$ is sufficiently small for the near equilibrium condition to hold. This explains an important feature of the concurrence in Fig.~\ref{fermionneq} (a)  that, as $\Delta \mu$ increases, there is an initially increasing interval for $\mu_1<\omega$, while there is no such interval if $\mu_1>\omega$.

Combining the perspectives in the near equilibrium regime and the strong nonequilibrium regime, and interpolating the regimes in between the two extremes, we can roughly account for the major features of the concurrence and the coherence $\rho_{34}$ in the entire nonequilibrium regime. For instance,  the non-monotonic behavior of the concurrence for $\mu_1<\omega$ is explained by the fact that it first goes up in the near equilibrium regime and that it eventually approaches an asymptotic value lower than its equilibrium value, which implies that it has to go down somewhere in the middle (assuming that it is a continuous function). This approach of interpolation cannot explain more specific features in the moderate nonequilibrium regime though. Fortunately, in our case, there seems to be nothing bizarre in between.

The concurrence can also be investigated from the perspective of the competition between the coherence $|w|$ (in the bare-state representation) and the population $\sqrt{ad}$ ($a=\rho_{11}$ and $d=\rho_{22}$). It turns out that the qualitative features of the concurrence in relation to $\Delta \mu$ is captured by those of the coherence $|w|$, as is evident by comparing Fig.~\ref{fermionneq} (a) and (c). For instance, $|w|$ is also a non-monotonic function of $\Delta \mu$ for $\mu_1<\omega$. The population term $\sqrt{ad}$ is a monotonic function of $\Delta\mu$, which did not alter the qualitative features of the coherence $|w|$. Hence, as far as the major features of the concurrence is concerned, the coherence $|w|$ comes out a winner in the competition with the population term $\sqrt{ad}$.

We also remark that, the entanglement in the fermion bath case is about one order larger than in the boson bath case in the parameter regimes considered, as can be seen by comparing Fig.~\ref{bosonneq}(a) with Fig.~\ref{fermionneq}(a). This seems to suggest that fermion baths may have an advantage in enhancing entanglement in the nonequilibrium setting compared to boson baths. However, one needs to be careful with the interpretation of this result. Two different nonequilibrium conditions are used in the boson bath case and the fermion bath case, namely, the temperature difference accompanied by energy exchange for the boson bath case and the chemical potential difference accompanied by particle exchange for the fermion bath case. Hence, the difference in the results may arise from two distinct factors, that is, different nonequilibrium mechanisms (energy exchange versus particle exchange) and different statistics (Bose-Einstein statistics versus Fermi-Dirac statistics). Our investigation on this issue with analytical solutions (applicable even if $\mu\neq 0$ for boson baths) as well as numerical calculations suggests that both factors played a role. We find that the exchange of particle (induced by the chemical potential difference) between the system and reservoirs generally has a beneficial effect on enhancing the steady-state entanglement than the exchange of energy (induced by the temperature difference) when the statistics of the baths is fixed. As to the statistics, one way it played a role is that the chemical potential of boson baths can only be negative while that of fermion baths has no such restriction.

\section{Analytical solution and entanglement phase diagrams for asymmetric qubits}\label{phased}

In previous sections we investigated the steady-state entanglement and coherence of the symmetric qubit system $(\omega_1=\omega_2)$ in both equilibrium and nonequilibrium settings. In this section we study the more general scenario when the two qubits  have an energy detuning ($\omega_1\neq \omega_2$).

We have also obtained the analytical solution for the asymmetric qubit case, which turns out to be generalizable from that for the symmetric qubit case by a simple map. By introducing
\begin{equation}\label{GNbartilde}
\bar{\mathcal{N}}_{\pm}=\bar{N}_{\pm}\pm \widetilde{N}_{\pm}\cos\theta, \quad \widetilde{\mathcal{N}}_{\pm}=\widetilde{N}_{\pm}\sin\theta,
\end{equation}
the asymmetric qubit solution can be obtained by replacing $\bar{N}_{\pm}$ and $\widetilde{N}_{\pm}$ in the symmetric qubit solution with $\bar{\mathcal{N}}_{\pm}$ and $\widetilde{\mathcal{N}}_{\pm}$, respectively. (Recall the definitions $\bar{N}_{\pm}=(N_1^{\pm}+N_2^{\pm})/2$ and $\widetilde{N}_{\pm}=(N_1^{\pm}-N_2^{\pm})/2$.) More explicitly, for asymmetric qubits coupled to boson baths, the steady-state solution reads
\begin{eqnarray}
\rho_{11}&=&\frac{1}{\mathcal{N}}\left[\bar{\mathcal{N}}_+\bar{\mathcal{N}}_--r_1r_2R\right],\label{APopulationBB11}\\
\rho_{22}&=&\frac{1}{\mathcal{N}}\left[(1+\bar{\mathcal{N}}_+)(1+\bar{\mathcal{N}}_-)-s_1s_2R\right],\\
\rho_{33}&=&\frac{1}{\mathcal{N}}\left[\bar{\mathcal{N}}_+(1+\bar{\mathcal{N}}_-)+s_1r_2R\right],\\
\rho_{44}&=&\frac{1}{\mathcal{N}}\left[\bar{\mathcal{N}}_-(1+\bar{\mathcal{N}}_+)+s_2r_1R\right],\label{APopulationBB44}\\
\rho_{34}&=&-\frac{1}{\mathcal{N}}\left[\frac{\widetilde{\mathcal{N}}_+(1+2\bar{\mathcal{N}}_-)+\widetilde{\mathcal{N}}_-(1+2\bar{\mathcal{N}}_+)}{2(1+\bar{\mathcal{N}}_++\bar{\mathcal{N}}_-) +i\Omega'}\right],
\end{eqnarray}
where $\mathcal{N}$ is the normalization factor and
\begin{eqnarray}
r_1&=&\widetilde{\mathcal{N}}_++\widetilde{\mathcal{N}}_-(1+2\bar{\mathcal{N}}_++2\bar{\mathcal{N}}_-),\\
r_2&=&\widetilde{\mathcal{N}}_-+\widetilde{\mathcal{N}}_+(1+2\bar{\mathcal{N}}_++2\bar{\mathcal{N}}_-),\\
s_1&=&\widetilde{\mathcal{N}}_+-\widetilde{\mathcal{N}}_-(3+2\bar{\mathcal{N}}_++2\bar{\mathcal{N}}_-),\\
s_2&=&\widetilde{\mathcal{N}}_--\widetilde{\mathcal{N}}_+(3+2\bar{\mathcal{N}}_++2\bar{\mathcal{N}}_-),\\
R&=&\frac{1}{4(1+\bar{\mathcal{N}}_++\bar{\mathcal{N}}_-)^2+\Omega'^2}.\label{AR}
\end{eqnarray}
For asymmetric qubits coupled to fermion baths, the steady-state solution is given by
\begin{eqnarray}
\rho_{11}&=&\bar{\mathcal{N}}_+\bar{\mathcal{N}}_--\widetilde{R},\label{APopulationFBG11}\\
\rho_{22}&=&(1-\bar{\mathcal{N}}_+)(1-\bar{\mathcal{N}}_-)-\widetilde{R},\\
\rho_{33}&=&\bar{\mathcal{N}}_+(1-\bar{\mathcal{N}}_-)+\widetilde{R},\\
\rho_{44}&=&\bar{\mathcal{N}}_-(1-\bar{\mathcal{N}}_+)+\widetilde{R},\label{PopulationFBG44}\\
\rho_{34}&=&-\frac{\widetilde{\mathcal{N}}_++\widetilde{\mathcal{N}}_-}{2+i\Omega'},
\end{eqnarray}
where
\begin{equation}\label{AExpR}
\widetilde{R}=\frac{(\widetilde{\mathcal{N}}_++\widetilde{\mathcal{N}}_-)^2}{4+\Omega'^2}
\end{equation}
and $\Omega'=\Omega/J$. The coherence in the bare-state representation can be calculated with $w=\sin\theta(\rho_{33}-\rho_{44})/2+\cos^{2}(\theta/2)\rho_{34}-\sin^{2}(\theta/2)\rho_{43}$ and the concurrence is given by $\mathcal{C}=2\max(0,|w|-\sqrt{\rho_{11}\rho_{22}})$.

The symmetric qubit solutions in Eqs.~(\ref{PopulationBB11})-(\ref{NormBB}) for boson baths and Eqs.~(\ref{PopulationFBG11})-(\ref{ExpR}) for fermion baths are special cases of the above general solution given that $\bar{\mathcal{N}}_{\pm}$ and $\widetilde{\mathcal{N}}_{\pm}$ reduce to $\bar{N}_{\pm}$ and $\widetilde{N}_{\pm}$, respectively, when $\omega_1=\omega_2$ ($\theta=\pi/2$). The structure of the maps in Eq.~(\ref{GNbartilde}) suggests a close connection between the detuning in the two qubits (manifested in the mixing angle $\theta$) and the nonequilibrium conditions (indicated by $\widetilde{N}_{\pm}$). Our numerical results support such a connection.

\begin{figure}[tbp]
\centering
\includegraphics[width=8cm]{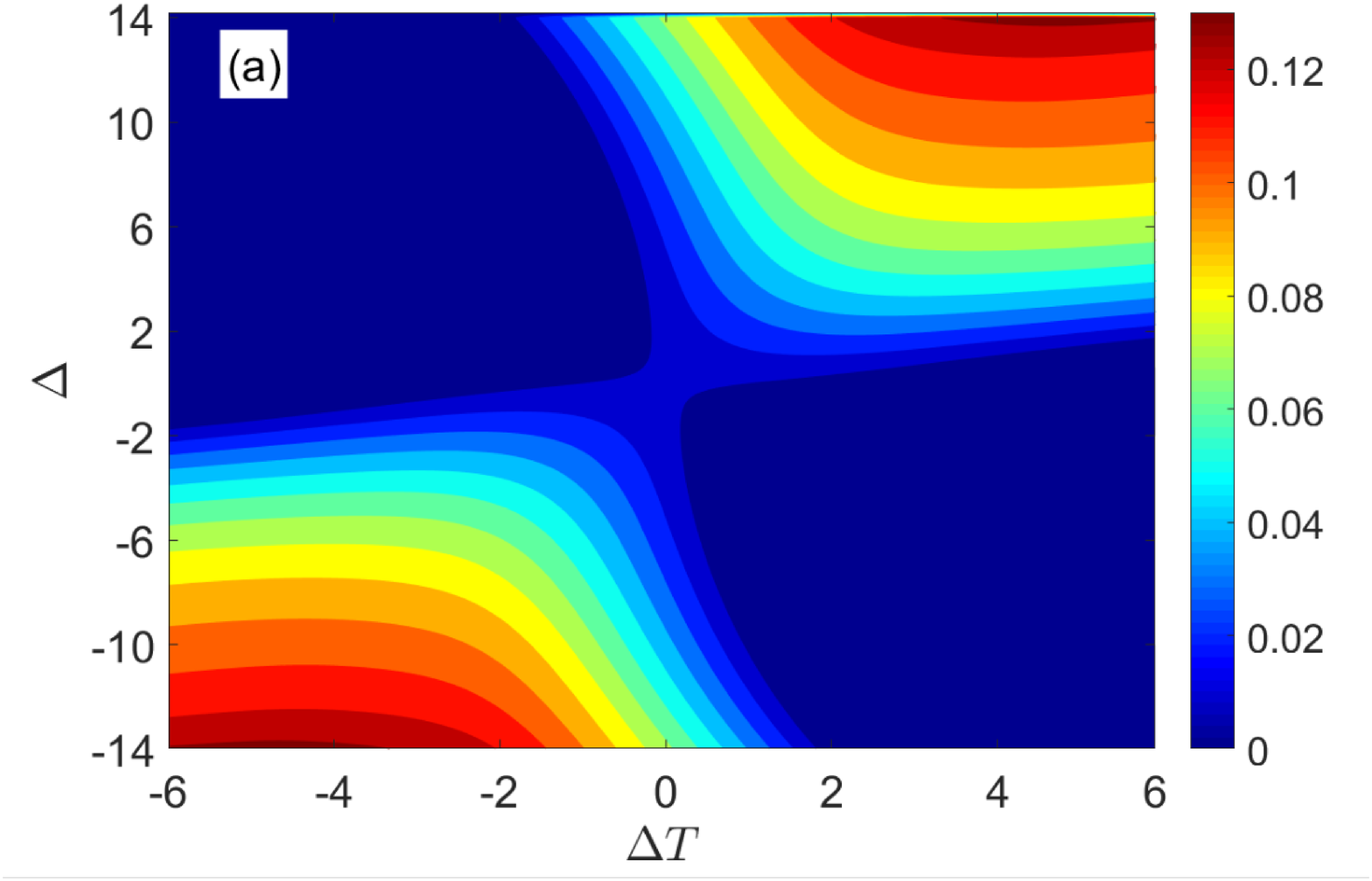}
\includegraphics[width=8cm]{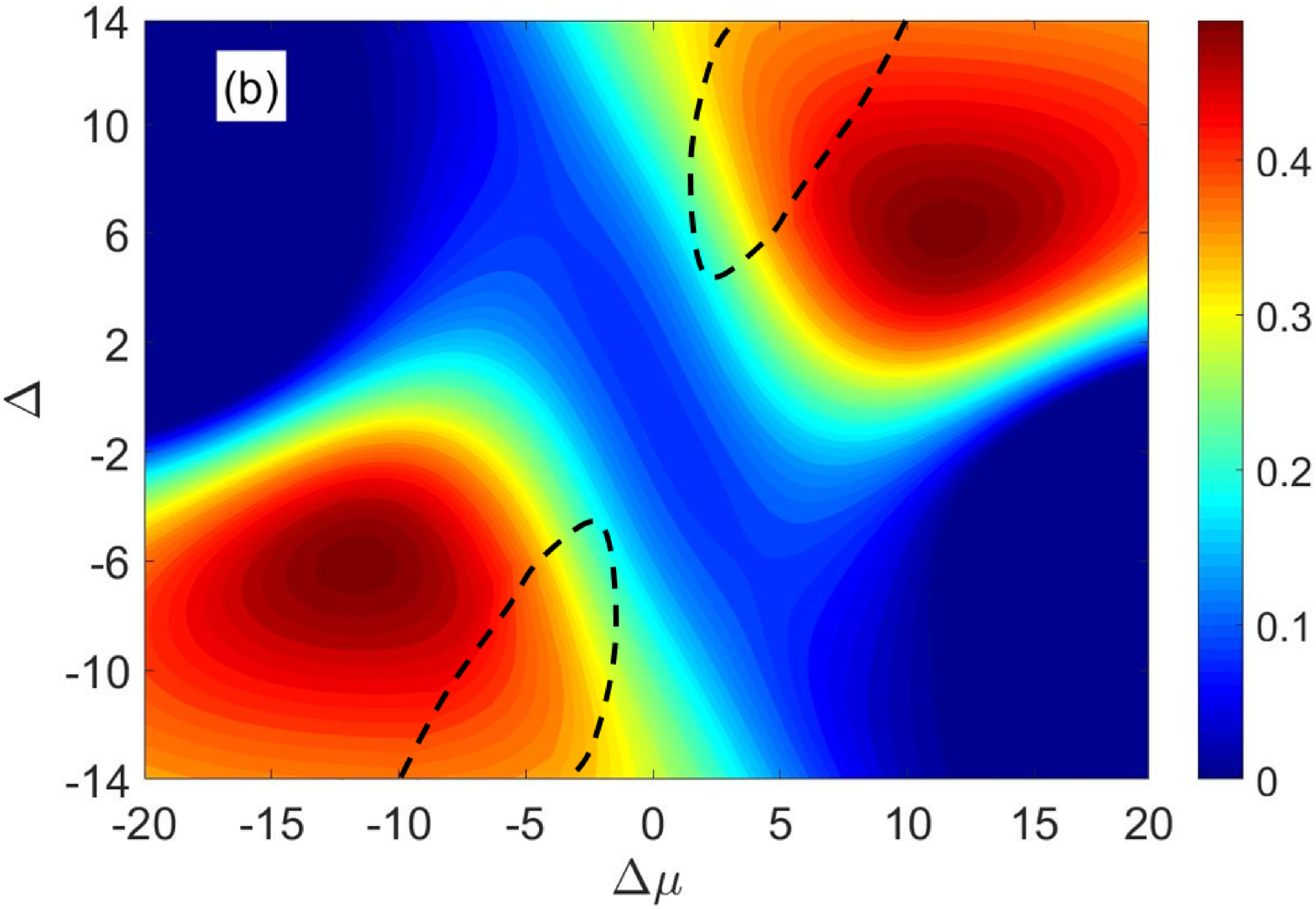}
\caption{(Color online)The entanglement phase diagrams of the system of the two coupled qubits immersed in boson baths (a) and fermion baths (b). The parameters are set as $J_1=J_2=1$, $\lambda=6$, $\bar{\omega}=10$ in (a) and (b); $\bar{T}=3$ in (a); $T_1=T_2=1.5$, $\bar{\mu}=4$ in (b).}
\label{phase}
\end{figure}

The numerical results are presented in the entanglement phase diagrams in Fig.~\ref{phase} (a) for boson reservoirs and  Fig.~\ref{phase} (b) for fermion reservoirs. The entanglement phase diagrams are contour plots of the steady-state entanglement, quantified by concurrence, as functions of the nonequilibrium condition ($\Delta T$ for boson reservoirs and $\Delta \mu$ for fermion reservoirs) and the energy detuning $\Delta=\omega_1-\omega_2$ of the two qubits.

The symmetry in the setup of the qubits and the baths (switching qubit one and bath one simultaneously with qubit two and bath two, respectively, would not change the physics) is important for our discussions in this section. To highlight this symmetry, we adopt a symmetric way of varying the parameters, which is different from that used in the previous sections but does not change the essential physics. More specifically, in the entanglement phase diagrams, we keep fixed $\bar{\omega}=(\omega_1+\omega_2)/2$ for the two qubits, $\bar{T}=(T_1+T_2)/2$ for the two boson baths, and $\bar{\mu}=(\mu_1+\mu_2)/2$ for the two fermion baths, while varying $\Delta=\omega_1-\omega_2$, $\Delta T=T_2-T_1$ and $\Delta\mu=\mu_2-\mu_1$, respectively. The ranges for the varying parameters are restricted by $|\Delta|<\sqrt{4\bar{\omega}^2-\lambda^2}$ according to the rotating wave approximation, $|\Delta T|\leq 2\bar{T}$ to ensure that the temperatures of both boson baths are non-negative, and unrestricted $\Delta\mu$ as the chemical potentials of the fermion baths can be positive or negative. As one can see in Fig.~\ref{phase} (a) and (b), the entanglement phase diagrams are symmetric with respect to the origin, reflecting the symmetry in the setup of the qubits and the baths.

In  the entanglement phase diagram for the boson bath case in Fig.~\ref{phase} (a), the entanglement has higher values in the top-right and bottom-left corners, where its value can reach about 0.12, roughly ten times that in the equilibrium case (for $\lambda=6$ and $T=3$). In the top-right corner, $\Delta>0$ and $\Delta T>0$, that is, $\omega_1>\omega_2$ and $T_1<T_2$ (note that $\Delta=\omega_1-\omega_2$ while $\Delta T=T_2-T_1$). Accordingly, in the bottom-left corner $\omega_1<\omega_2$ and $T_1>T_2$. This suggests that, to enhance the steady-state entanglement of the system by exploiting the nonequilibrium condition and the detuning of the two qubits, it is advantageous to couple the qubit with a higher frequency to the boson bath with a lower temperature.

For the fermion bath case, the entanglement phase diagram in Fig.~\ref{phase} (b) shows that the concurrence has a maximum within quadrant I and another maximum in quadrant III due to the symmetric setup. The maximum concurrence achieved is close to $0.5$, about five times that of the equilibrium fermion bath case (at $\mu=4$) and also about five times the value in the phase diagram of the boson bath base. In addition, coupling the qubit with a higher frequency to the fermion bath with a lower chemical potential also has a beneficial effect on promoting the steady-state entanglement.

The above observation that high entanglement is achieved for asymmetric qubits $\omega_2<\omega_1$ when $T_2>T_1$ (boson baths) or $\mu_2>\mu_1$ (fermion baths) suggests a compensation mechanism between the detuning of the two qubits and the nonequilibrium condition of the two baths. A possible explanation is that the difference of the temperature or chemical potential of the two baths may strengthen the effective coupling between the two detuned qubits, leading to elevated entanglement between them.

We remark that there are two regions in Fig.~\ref{phase} (b) with their boundaries marked out with dashed black lines (the locations of the boundaries are not intended to be exact). In these two regions there are minor violations of the positivity of the density matrix ($\rho_{11}$ has negative values of the order $10^{-4}$), suggesting that the conditions in these parameter regimes are outside the validity range of the Bloch-Redfield equation.  The phase diagram within these regions are obtained by taking only the real part of $\sqrt{ad}$ in the calculation of the concurrence, producing a smooth transition across the boundaries of these regions. A more rigorous treatment may involve comparing with steady-state solutions to the exact dynamics of the system so that the validity regime of the Bloch-Redfield equation in this case may be quantified, which is reserved for future work. We note that all other figures in previous sections were obtained within parameter regimes that guarantee the positivity of the density matrix.

{\section{Effect of spectral densities and connection to energy current}\label{SpectralEnergy}}

{The results in previous sections were restricted to balanced and frequency-independent spectral densities, namely, $J_1(\delta/2\pm\Omega/2)=J_2(\delta/2\pm\Omega/2)=J$. It is worthwhile to investigate how more complex spectral densities may affect those findings. In addition, the nonequilibrium nature of the system is also signified by the nonvanishing energy current at steady state, characterizing the transport features. It is of interest to explore how the energy current is connected to the nonequilibrium condition and the steady-state coherence and entanglement. An in-depth study of these problems entails extensive exposition that goes beyond the scope of the present paper. In this section we
present some preliminary results on these questions and leave more systematic investigations for future work.}

{\subsection{Reservoirs with the Ohmic spectrum}\label{OhmicSpectrum}}

\begin{figure}[tbp]
\centering
\includegraphics[width=8cm]{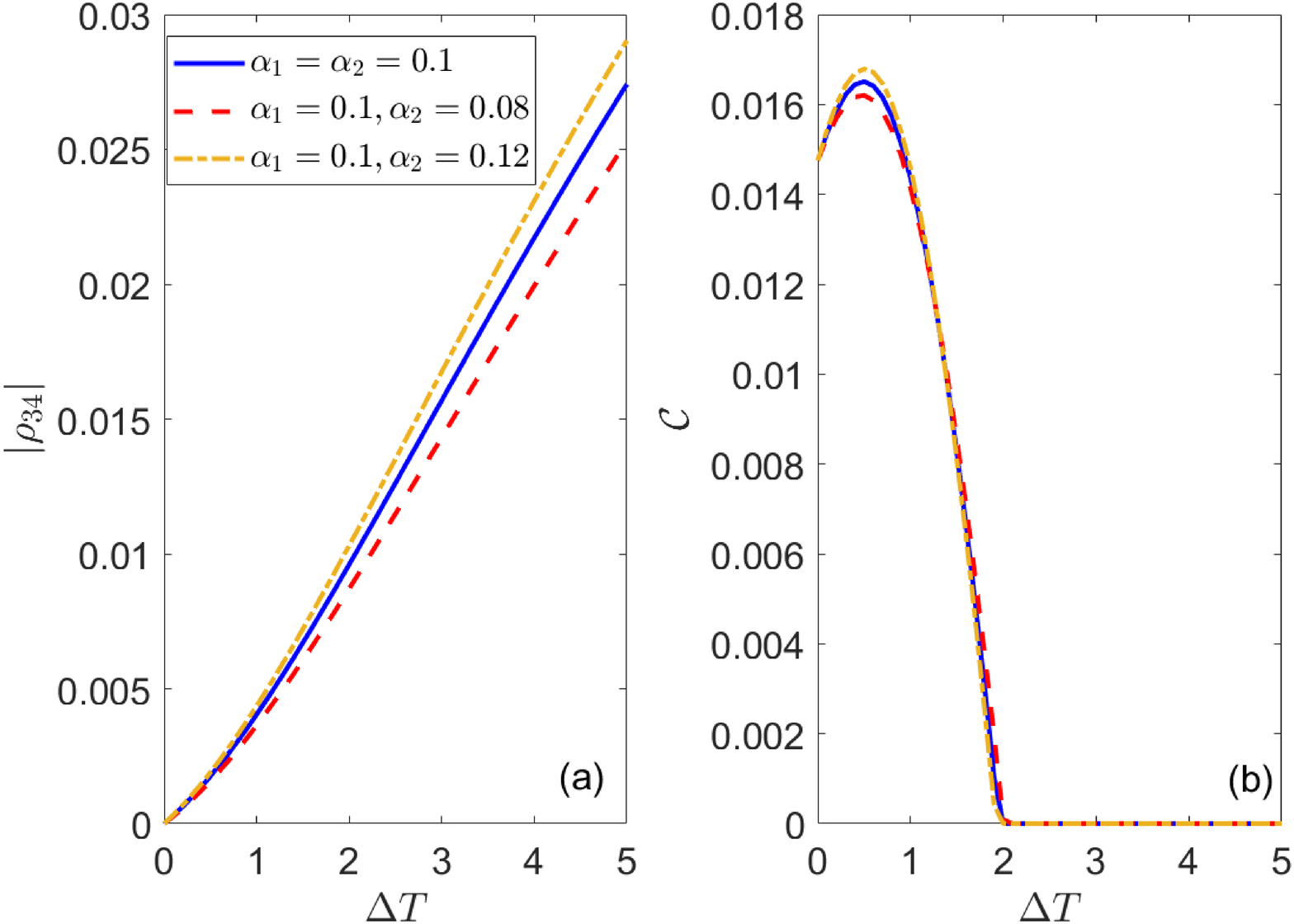}
\includegraphics[width=8cm]{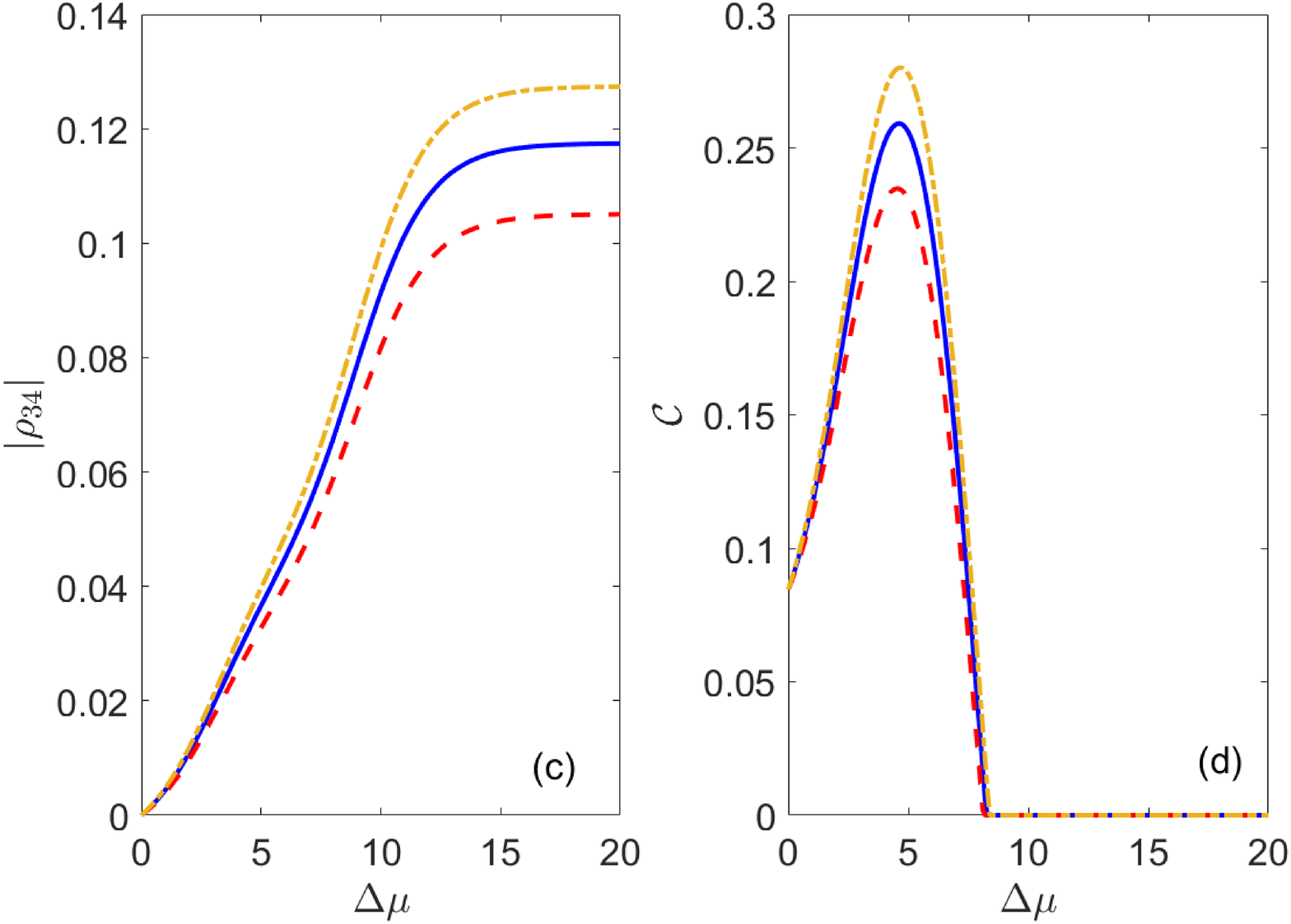}
{\caption{(Color online) The steady-state coherence and entanglement for boson reservoirs (a,b) and fermion reservoirs (c,d). The parameters are set as $\omega_1=\omega_2=10, \omega_{c1}=\omega_{c2}=40$; (a,b) $T_1=2, T_2=T_1+\Delta T$; (c,d) $T_1=T_2=1.5,\mu_1=4,\mu_2=\mu_1+\Delta\mu$.}\label{spectrum}}
\end{figure}

{To investigate the effect of spectral densities, we consider reservoirs with the Ohmic spectrum~\cite{legget}
\begin{equation}
\begin{split}
J_1(\omega)&=\alpha_1\omega\exp(-\omega/\omega_{c1}),\\
J_2(\omega)&=\alpha_2\omega\exp(-\omega/\omega_{c2}),
\end{split}
\end{equation}
where $\alpha_1$ and $\alpha_2$ are the dissipation coefficients and $\omega_{c1}$ and
$\omega_{c2}$ denote the cutoff frequencies. Note that, in order to guarantee the validity of the Markovian approximation, the parameters in the spectral densities need to conform with the condition $\{J_1(\omega),J_2(\omega)\}\ll\{\omega_1,\omega_2,\lambda\}$, especially in the vicinity of $\omega=\delta/2\pm\Omega/2$.}

{The steady-state coherence and entanglement are plotted in Fig.~\ref{spectrum} (a) and (b) for boson reservoirs with a temperature difference and Fig.~\ref{spectrum} (c) and (d) for fermion reservoirs with a chemical potential difference. Compared to the balanced and frequency-independent cases shown in Figs.~\ref{bosonneq} and \ref{fermionneq}, the results in Fig.~\ref{spectrum}, obtained with the Ohmic spectrum, are not so much qualitatively different.  However, these results do suggest that unbalanced couplings to the reservoirs in a particular manner may enhance the steady-state coherence and entanglement. More specifically, a stronger coupling between the system and the reservoir with a higher temperature (boson reservoirs) or chemical potential (fermion reservoirs) tends to increase the steady-state coherence and entanglement. This is reflected in Fig.~\ref{spectrum} by the fact that the steady-state coherence and entanglement have higher (lower) values for $\alpha_2>\alpha_1$ ($\alpha_2<\alpha_1$) than for $\alpha_2=\alpha_1$, when $T_2> T_1$ or $\mu_2> \mu_1$.}

{\subsection{Connection to the energy current}}

{To investigate the energy current at the nonequilibrium steady state, we reorganize the dissipators in the master equation in Eq.~(\ref{master}) according to the labels of each individual reservoir. More specifically, the dissipator can be rewritten as $D_0[\rho]+D_s[\rho]=D_1[\rho]+D_2[\rho]$, where $D_1[\rho]$ ($D_2[\rho]$) is the dissipator associated with the reservoir in contact with qubit $1$ (qubit $2$). Here $D_i[\rho]=\mathcal{N}_i[\rho]+\mathcal{S}_i[\rho]$ ($i=1, 2$), where the expressions of $\mathcal{N}_i[\rho]$ and $\mathcal{S}_i[\rho]$ are given in Eqs.~(\ref{seque}) and (\ref{noseque}). The energy current from the $i$-th reservoir to the system at the steady state is given by $I_i=\dot{Q}_i={\rm Tr}\{D_i[\rho_{ss}]H_s\}$ ($i=1, 2$)~\cite{current1,current2,current3,current4}. Energy conservation at the steady state dictates that $I_1+I_2=0$. Thus, without loss of generality, we only focus on the energy current $I_2$.}

\begin{figure}[tbp]
\centering
\includegraphics[width=8cm]{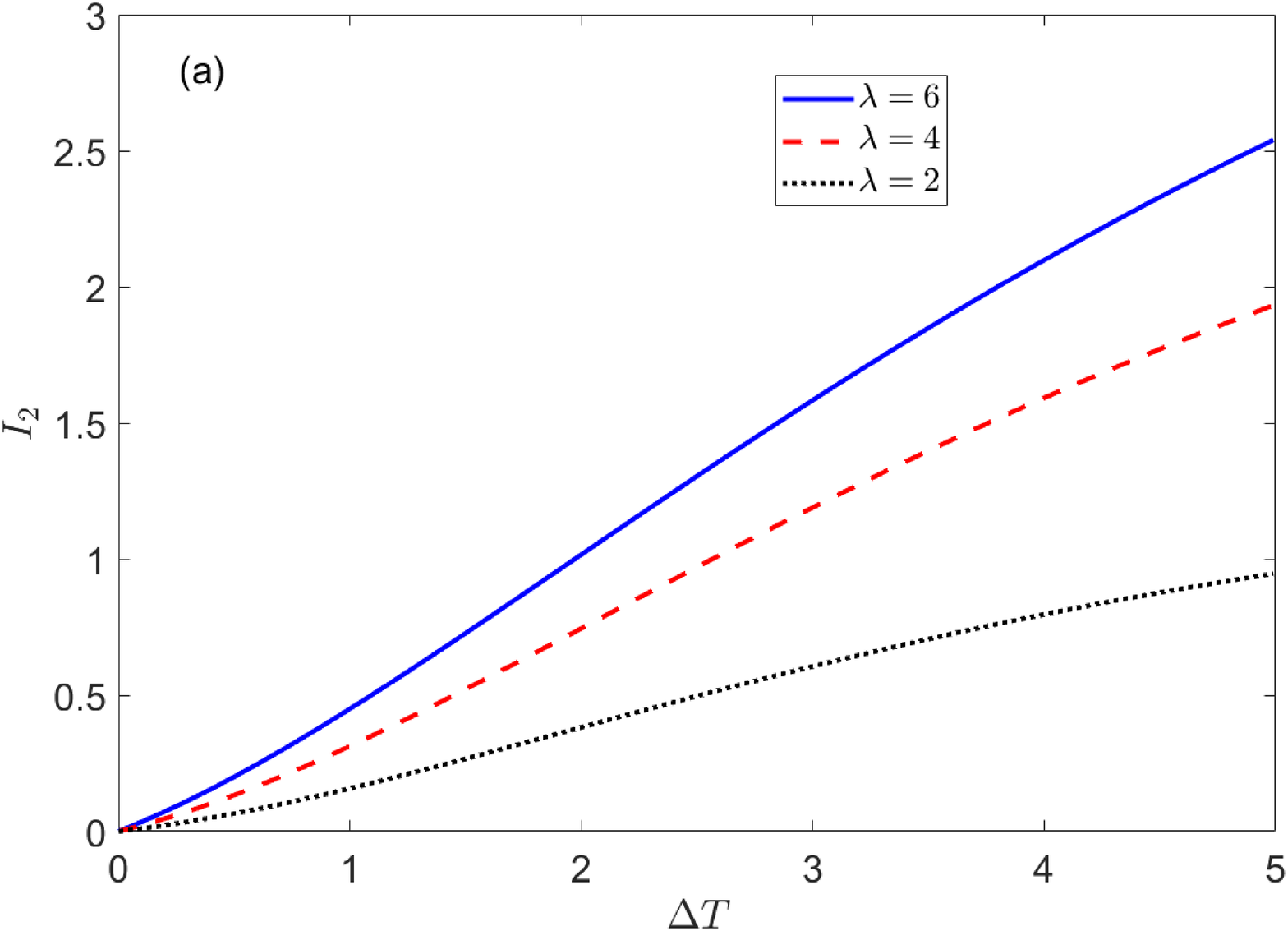}
\includegraphics[width=8cm]{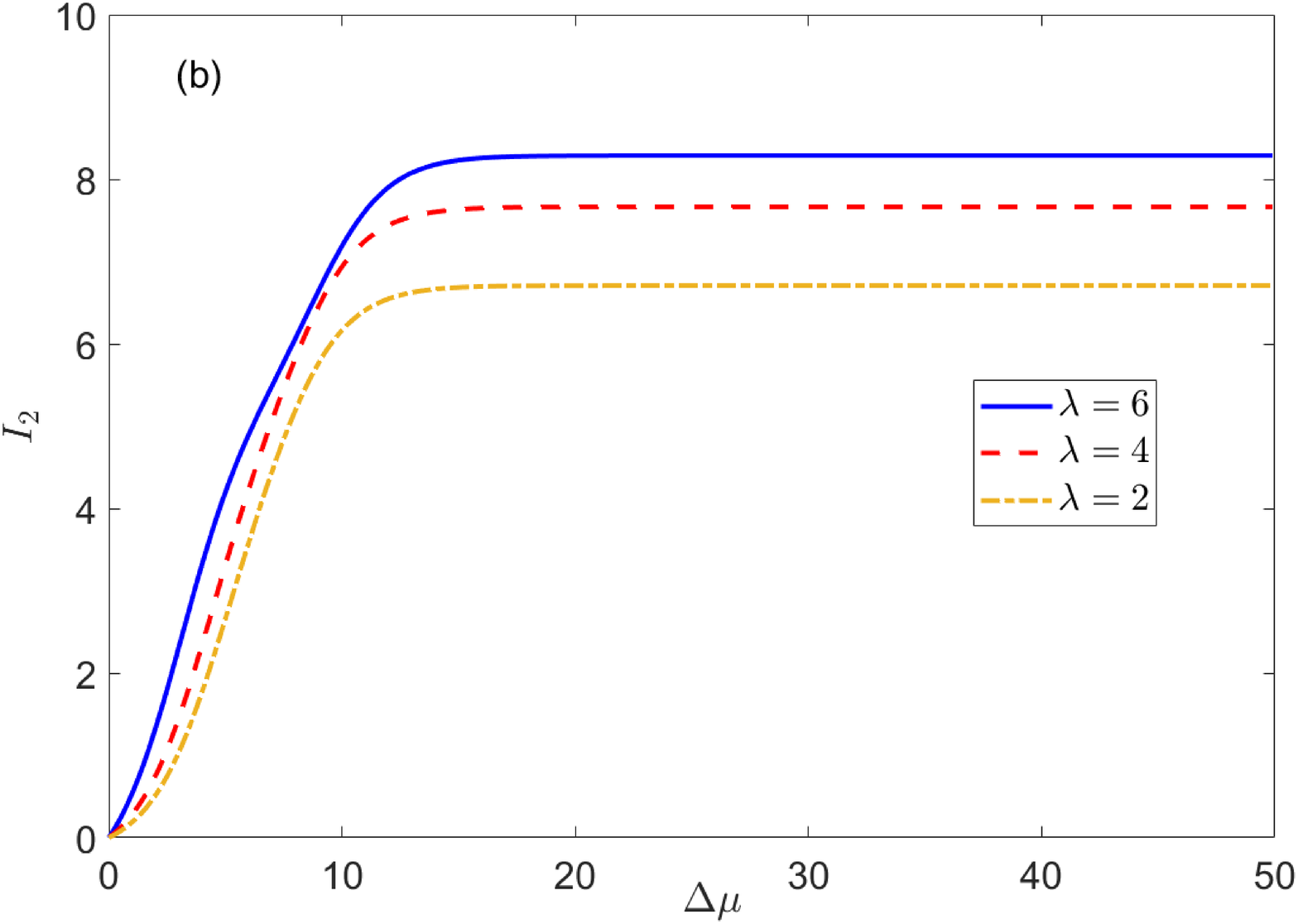}
{\caption{(Color online) The energy current in the steady-state for boson reservoirs (a) and fermion reservoirs (b) for different inter-qubit coupling strengthes.  The parameters are set as $\omega_1=\omega_2=10, J_1=J_2=1$; (a) $T_1=2, T_2=T_1+\Delta T$;  (b) $T_1=T_2=1.5,\mu_1=4,\mu_2=\mu_1+\Delta\mu$.}\label{current}}
\end{figure}

{In Fig.~\ref{current}, for different inter-qubit coupling strengths, we plotted the energy current as a function of the nonequilibrium condition (temperature difference for boson reservoirs in Fig.~\ref{current}(a) and chemical potential difference for fermion reservoirs in Fig.~\ref{current}(b)). As one can see, $I_2 > 0$ when $T_2>T_1$ (boson reservoirs) or $\mu_2>\mu_1$ (fermion reservoirs), indicating that an energy current flows from the reservoir with a higher temperature or chemical potential to that with a lower one through the coupled qubit system, as expected. In both cases, the energy current increases with the nonequilibrium condition and the inter-qubit coupling strength. Further comparison with Fig.~\ref{bosonneq}(b) and Fig.~\ref{fermionneq}(b) reveals that, for both boson and fermion reservoirs, the energy current displays monotonic behaviors with respect to the nonequilibrium condition, similar to the steady-state coherence. The resemblance is also evident in the feature of saturation at large chemical potential difference for the fermion reservoir case, as seen in Fig.~\ref{fermionneq}(b) and Fig.~\ref{current}(b). This suggests a close connection between the energy current and the steady-state coherence (in the energy representation) that both originate from the nonequilibrium condition \cite{coh2,coh6}. In contrast, there does not seem to be a simple correlation between the energy current and the steady-state entanglement. This may be partly due to the fact that the energy current is partitioned to support both steady-state coherence and populations \cite{coh6}, while the concurrence quantifying entanglement in our case is a result of the competition between the coherence and the populations as discussed in Sec.~\ref{ent}.}

\section{Conclusion}
\label{conclusion}

In this paper, we studied the steady-state entanglement and coherence of two coupled qubits each embedded in a local boson or fermion reservoir, using the Bloch-Redfield master equation beyond the secular approximation. We obtained general analytical solutions to the steady state of the master equation, which, combined with numerical results, allowed us to explore in detail the behaviors of entanglement and coherence at the steady state. Most features of the entanglement and coherence can be accounted for by interpolating their asymptotic behaviors in extreme conditions (e.g., the near equilibrium regime and the strong nonequilibrium regime) and from the perspective that concurrence is determined by the competition between coherence and population in the bare-state representation.

In the equilibrium situation, we found that the entanglement varies non-monotonically with the temperature or chemical potential and becomes significantly diminished if the inter-qubit coupling is too weak, while the coherence in the eigen-state representation vanishes due to decoherence as expected. In the nonequilibrium situation, there is non-vanishing steady-state coherence in the eigen-state representation that grows monotonically with the nonequilibrium condition (temperature difference or chemical potential difference), while the steady-state entanglement in general behaves non-monotonically with the nonequilibrium condition. When the base temperature or chemical potential is high enough, the steady-state entanglement and coherence, however, become strongly suppressed irrespective of the strength of the nonequilibrium condition. We also demonstrated in the entanglement phase diagrams that combining the nonequilibrium condition with the detuning of the two qubits in a compensatory way (coupling the qubit with higher frequency to the reservoir with lower temperature or chemical potential) can boost the steady-state entanglement roughly $5\sim 10$ times that of the corresponding equilibrium symmetric qubit case. In addition, there is sizable improvement in entanglement when the qubits are coupled to fermion reservoirs exchanging particles with the system compared to boson reservoirs exchanging energy with the system under similar conditions.

Our study suggests some viable strategies that may be used to benefit the optimization of the steady-state entanglement in the coupled qubit system, which include the following: use sufficiently strong inter-qubit coupling; maintain the temperature or chemical potential of one bath at a relatively low level; keep the temperature difference or chemical potential difference of the two baths at a moderate level; couple the qubit with a higher frequency to the bath with a lower temperature or chemical potential; {implement a stronger coupling between the system and the reservoir with a higher temperature or chemical potential;} couple the qubits to fermion reservoirs exchanging particles with the system over boson reservoirs exchanging energy with the system. These strategies are intended to be used as general guidelines which need to be supplemented with more detailed analysis of the steady-state entanglement in relation to the inter-qubit coupling strength, the detuning of the two qubits, the nonequilibrium condition etc as done in this work, in order to achieve optimized results of enhanced steady-state entanglement.

There is a possibility that our results on the nonequilibrium enhanced steady-state entanglement and coherence may be experimentally tested in the foreseeable future. For the case of boson reservoirs, the coupled-qubit system can be realized by superconducting charge qubits~\cite{yap}, and the nonequilibrium condition indicated by the temperature difference of the reservoirs can be adjusted by tuning diluted magnetic refrigerators. The nonequilibrium condition may also be created by coupling the qubits to reservoirs with different coupling strengths. On the other hand, the fermion reservoir case can be experimentally realized using hybrid circuit-QED~\cite{QED}, in which semiconducting quantum dots define the qubits that are coupled to the leads serving as electron reservoirs~\cite{QED2,ahk}. The nonequilibrium condition characterized by the chemical potential difference can be created by the bias voltage of the leads. Furthermore, quantum state tomography has been widely applied in reconstructing the density matrix of quantum systems. In particular, the schemes to reconstruct the density matrix of the two-qubit system have been proposed~\cite{SF,NR}. The steady-state entanglement and coherence can thus be obtained from the experimentally reconstructed steady-state density matrix. Therefore, there is a good chance the theoretical and numerical results presented in this paper can be tested against experiments in the near future.

The approach in this study and some of the general guidelines proposed for the coupled qubit system may be extended to more general settings to optimize the nonequilibrium steady-state entanglement and coherence, which may have potential applications in quantum communication and the design of quantum devices working in noisy nonequilibrium environments. In future work, we intend to quantify the range of validity of the Bloch-Redfield master equation used in this study and compare the results here with those based on exact dynamics of the system with the non-Markovian effects fully taken into account. {A more systematic investigation on the influence of spectral densities in the non-Markovian regime as well as the connection to the energy current will also be considered.}

\begin{acknowledgments}
Z. W. is supported by NSFC (Grant No. 11875011), and Educational Commission of Jilin Province of China (Grant No. JJKH20190266KJ). W. W. is supported by Chinese Academy of Sciences (Project No. YJKYYQ 20180038) and Chinese Ministry of Science and Technology (Project No. 2016YFA0203200). J. W. acknowledges the funding support from NSF (Grant No. NSF-PHY-76066 and NSF-CHE-1808474).
\end{acknowledgments}

\appendix
\addcontentsline{toc}{section}{Appendices}\markboth{APPENDICES}{}
\begin{subappendices}

\section{The quantum master equation in terms of the density matrix elements}
\label{appendix1}

The operator form of the quantum master equation in Eqs.~(\ref{master})-(\ref{noseque}) in the main text can be written in terms of the density matrix elements in the eigen-state representation as follows:
\begin{equation}
\frac{d}{dt}\rho_{ij}=\sum_{lk}\mathcal{M}_{ij}^{lk}\rho_{lk},
\end{equation}
where
\begin{equation}
\mathcal{M}_{11}^{11}=-2[\sin^{2}\frac{\theta}{2}
(\Gamma_{1}^{-}+\Gamma_{2}^{+})
+\cos^{2}\frac{\theta}{2}(\Gamma_{1}^{+}
+\Gamma_{2}^{-})],
\end{equation}
\begin{equation}
\mathcal{M}_{11}^{33}=2(\sin^{2}\frac{\theta}{2}\gamma_{1}
^{-}+\cos^{2}\frac{\theta}{2}\gamma_{2}^{-}),
\end{equation}
\begin{equation}
\mathcal{M}_{11}^{44}=2(\cos^{2}\frac{\theta}{2}\gamma_{1}^{+}
+\sin^{2}\frac{\theta}{2}
\gamma_{2}^{+}),
\end{equation}
\begin{equation}
\mathcal{M}_{11}^{34}=\mathcal{M}_{11}^{43}=\frac{1}{2}\sin\theta
(\gamma_{1}^{+}+\gamma_{1}^{-}-\gamma_{2}^{+}-\gamma_{2}^{-}),
\end{equation}
\begin{equation}
\mathcal{M}_{22}^{22}=-2[\sin^{2}\frac{\theta}{2}(\gamma_{1}^{-}
+\gamma_{2}^{+})+\cos^{2}\frac{\theta}{2}(\gamma_{1}^{+}+\gamma_{2}^{-})],
\end{equation}
\begin{equation}
\mathcal{M}_{22}^{33}=2(\cos^{2}\frac{\theta}{2}\Gamma_{1}^{+}+
\sin^{2}\frac{\theta}{2}\Gamma_{2}^{+}),
\end{equation}
\begin{equation}
\mathcal{M}_{22}^{44}=2(\cos^{2}\frac{\theta}{2}\Gamma_{2}^{-}+
\sin^{2}\frac{\theta}{2}\Gamma_{1}^{-})
\end{equation}
\begin{equation}
\mathcal{M}_{22}^{34}=\mathcal{M}_{22}^{43}=-\frac{1}{2}\sin\theta
(\Gamma_{1}^{+}+\Gamma_{1}^{-}-\Gamma_{2}^{+}-\Gamma_{2}^{-}),
\end{equation}
\begin{equation}
\mathcal{M}_{33}^{11}=2(\sin^{2}\frac{\theta}{2}\Gamma_{1}^{-}
+\cos^{2}\frac{\theta}{2}\Gamma_{2}^{-}),
\end{equation}
\begin{equation}
\mathcal{M}_{33}^{22}=2(\cos^{2}\frac{\theta}{2}\gamma_{1}^{+}
+\sin^{2}\frac{\theta}{2}\gamma_{2}^{+}),
\end{equation}
\begin{equation}
\mathcal{M}_{33}^{33}=-2[\sin^{2}\frac{\theta}{2}(\gamma_{1}^{-}
+\Gamma_{2}^{+})+\cos^{2}\frac{\theta}{2}(\gamma_{2}^{-}+\Gamma_{1}^{+})],
\end{equation}
\begin{equation}
\mathcal{M}_{33}^{34}=\mathcal{M}_{33}^{43}=-\frac{1}{2}\sin\theta
(\gamma_{1}^{+}-\Gamma_{1}^{-}-\gamma_{2}^{+}+\Gamma_{2}^{-}),
\end{equation}
\begin{equation}
\mathcal{M}_{44}^{11}=2(\cos^{2}\frac{\theta}{2}\Gamma_{1}^{+}+\sin^{2}\frac{\theta}{2}\Gamma_{2}^{+}
),
\end{equation}
\begin{equation}
\mathcal{M}_{44}^{22}=2(\sin^{2}\frac{\theta}{2}\gamma_{1}^{-}+\cos^{2}\frac{\theta}{2}\gamma_{2}^{-}
),
\end{equation}
\begin{equation}
\mathcal{M}_{44}^{44}=-2[\cos^{2}\frac{\theta}{2}(\gamma_{1}^{+}+\Gamma_{2}^{-})+\sin^{2}\frac{\theta}{2}(\gamma_{2}^{+}
+\Gamma_{1}^{-})],
\end{equation}
\begin{equation}
\mathcal{M}_{44}^{34}=\mathcal{M}_{44}^{43}=-\frac{1}{2}\sin\theta
(\gamma_{1}^{-}-\Gamma_{1}^{+}-\gamma_{2}^{-}+\Gamma_{2}^{+}),
\end{equation}
\begin{equation}
\mathcal{M}_{34}^{11}=\mathcal{M}_{43}^{11}=\frac{1}{2}\sin\theta(\Gamma_{1}^{+}+\Gamma_{1}^{-}
-\Gamma_{2}^{+}-\Gamma_{2}^{-}),
\end{equation}
\begin{equation}
\mathcal{M}_{34}^{22}=\mathcal{M}_{43}^{22}=-\frac{1}{2}\sin\theta(\gamma_{1}^{+}+\gamma_{1}^{-}
-\gamma_{2}^{+}-\gamma_{2}^{-}),
\end{equation}
\begin{equation}
\mathcal{M}_{34}^{33}=\mathcal{M}_{43}^{33}=-\frac{1}{2}\sin\theta(\gamma_{1}^{-}-\Gamma_{1}^{+}
-\gamma_{2}^{-}+\Gamma_{2}^{+}),
\end{equation}
\begin{equation}
\mathcal{M}_{34}^{44}=\mathcal{M}_{43}^{44}=-\frac{1}{2}\sin\theta(\gamma_{1}^{+}-\Gamma_{1}^{-}
-\gamma_{2}^{+}+\Gamma_{2}^{-}),
\end{equation}
\begin{equation}
\begin{split}
\mathcal{M}_{34}^{34}&=(\mathcal{M}_{43}^{43})^*=-\sin^{2}\frac{\theta}{2}(\gamma_{1}^{-}
+\Gamma_{1}^{-}+\gamma_{2}^{+}+\Gamma_{2}^{+})\\
&-\cos^{2}\frac{\theta}{2}(\gamma_{1}^{+}
+\Gamma_{1}^{+}+\gamma_{2}^{-}+\Gamma_{2}^{-})-i\Omega.
\end{split}
\end{equation}
\begin{equation}
\mathcal{M}_{11}^{22}=\mathcal{M}_{22}^{11}=\mathcal{M}_{33}^{44}=\mathcal{M}_{44}^{33}=\mathcal{M}_{34}^{43}=\mathcal{M}_{43}^{34}=0.
\end{equation}
Note that among all the off-diagonal elements of the density matrix, only $\rho_{34}$ and $\rho_{43}(=\rho_{34}^*)$ are relevant here, as the rest of them are decoupled and approach zero in the steady state.

\section{Method of solving the steady-state density matrix}
\label{appendix2}

Writing the relevant density matrix elements as a vector $|\rho\rangle=(\rho_{11},\rho_{22},\rho_{33},\rho_{44},\rho_{34},\rho_{43})^T$, the quantum master equation can be reformulated in the following vector-matrix form
\begin{equation}
\frac{d}{dt}|\rho\rangle=\mathcal{M}|\rho\rangle.
\end{equation}
The vector $|\rho\rangle$ can be partitioned into its population and coherence components, $|\rho\rangle=(\rho_p, \rho_c)^T$. Accordingly, the matrix $\mathcal{M}$ has the partitioned form
\begin{equation}
\mathcal{M}=\begin{bmatrix}
M_{pp}&M_{pc}\\
M_{cp}&M_{cc}
\end{bmatrix}.
\end{equation}
The steady state satisfies the equation $\mathcal{M}|\rho\rangle=0$, which reads
\begin{equation}
\left\{
\begin{array}{lcccl}
M_{pp}\rho_p+M_{pc}\rho_c&=&0\\
M_{cp}\rho_p+M_{cc}\rho_c&=&0
\end{array}.\right.
\end{equation}
The second equation yields $\rho_c=-M_{cc}^{-1}M_{cp}\rho_p$ (assuming $M_{cc}$ is invertible), which expresses $\rho_c$ in terms of $\rho_p$. Plugging it back into the first equation, one arrives at an equation involving only the population component:
\begin{equation}\label{FCME}
\mathcal{A}\,\rho_p=0,
\end{equation}
where
\begin{equation}
\mathcal{A}=M_{pp}-M_{pc}M_{cc}^{-1}M_{cp}.
\end{equation}

The equation $\mathcal{A}\,\rho_p=0$ has a reduced dimension compared to $\mathcal{M}|\rho\rangle=0$, which makes it easier to solve analytically. In our case here $\mathcal{A}$ is a $4\times 4$ matrix. Probability conservation ensures that each column of $\mathcal{A}$ adds up to zero, implying that its determinant is zero (its rank less than $n$). Typically, physical conditions may ensure that the rank of $\mathcal{A}$ is $n-1$ (in our case $3$), so that the equation $\mathcal{A}\,\rho_p=0$ has a unique solution up to normalization, which can be obtained as follows. Choose any row of $\mathcal{A}$, for instance, the first row, with the elements $(\mathcal{A}_{11}, \cdots, \mathcal{A}_{1i},\cdots, \mathcal{A}_{1n} )$. Then the $i$-th component of the solution $\rho_p$ is proportional to the cofactor (i.e., signed minor) of the matrix entry $\mathcal{A}_{1i}$. The proportionality factor is fixed by the normalization condition. With $\rho_p$ obtained, the coherence component is calculated with $\rho_c=-M_{cc}^{-1}M_{cp}\rho_p$. Thus one obtains all the relevant steady-state density matrix elements. More detailed explanations can be found in Ref.~\cite{coh6}.

\end{subappendices}

\end{document}